\documentclass[conference]{IEEEtran}
\IEEEoverridecommandlockouts

\usepackage{amsmath,amssymb,amsfonts}
\usepackage{amsthm}
\usepackage{graphicx}
\usepackage{textcomp}
\usepackage{xcolor}
\def\BibTeX{{\rm B\kern-.05em{\sc i\kern-.025em b}\kern-.08em
    T\kern-.1667em\lower.7ex\hbox{E}\kern-.125emX}}

\newcommand{\mc}{\mathcal}

\newcommand{\klone}{\textsc{klone}}
\newcommand{\kguard}{\textsc{kguard}}

\newcommand{\sub}[1]{_{_#1}}
\AtBeginDocument{%
  \providecommand\BibTeX{{%
    Bib\TeX}}}

\usepackage{tikz}
\usepackage[inline]{enumitem}

\usepackage{multirow}

\usepackage{biblatex}
\usepackage{hyperref}
\usepackage{algorithmicx}
\usepackage{algpseudocode}
\algdef{SE}[DOWHILE]{Do}{doWhile}{\algorithmdo}[1]{\algorithmwhile\ #1}%

\usepackage{enumitem}
\usepackage{overpic}
\usepackage[linesnumbered,ruled,vlined]{algorithm2e}
\usepackage{accsupp}

\usetikzlibrary{positioning, arrows.meta, shapes.geometric}
\newtheorem{definition}{Definition}[section]
\newtheorem{lemma}{Lemma}[section]
\newtheorem{proposition}{Proposition}[section]
\definecolor{arsenic}{rgb}{0.43, 0.5, 0.5}
\usepackage{marginnote}

\definecolor{darkgreen}{rgb}{0.0, 0.35, 0.09}

\DeclareMathOperator*{\argmax}{arg\,max}

\usepackage{cancel}
\usepackage{stackengine}
\usepackage{lipsum}
\usepackage{MnSymbol}
\usepackage{graphicx}

\newcommand*\mnote[3][0pt]{%
  \if l#2\reversemarginpar\def\pointer{\filledmedtriangleright}%
    \setlength\marginparsep{0.9cm}
    \def\stackalignment{r}\fi%
  \if r#2\normalmarginpar\def\pointer{\filledmedtriangleleft}%
    \setlength\marginparsep{0cm}
    \def\stackalignment{l}\fi%
  \marginpar{%
    \topinset{%
      \scalebox{1.5}{\textcolor{blue}{$\pointer$}}}{%
      \belowbaseline[-2\baselineskip-#1]{%
        \stackengine%
          {-5pt}%
          {\fcolorbox{blue}{white}{\parbox{1.5cm}%
            {\vspace{3pt}\raggedright#3}}}%
          {~\colorbox{white}{\sffamily Note}}%
          {O}%
          {l}%
          {F}%
          {F}%
          {S}%
        }%
      }{%
      3ex+#1}{%
      -2ex}%
  }%
}
\usepackage{listings}
\usepackage{xcolor}
\usepackage{subcaption}

\usepackage{tcolorbox}
\usepackage{xcolor}

\definecolor{lightgray}{RGB}{220,220,220}

\definecolor{verylightgray}{RGB}{240,240,240} 
\usepackage{booktabs}
\usepackage{pifont}

\newtcolorbox{graybox}{
    colback=verylightgray,
    sharp corners,
    boxrule=0pt,
    boxsep=0pt,
    left=4pt,
    right=4pt,
    top=4pt,
    bottom=4pt
}
\AtBeginDocument{%
  \providecommand\BibTeX{{%
    Bib\TeX}}}

\bibliography{ICDE/biblioKG_icde} 

  \makeatletter
\newcommand{\linebreakand}{%
 \end{@IEEEauthorhalign}
 \hfill\mbox{}\par
 \mbox{}\hfill\begin{@IEEEauthorhalign}
}
\makeatother

\usepackage{color, colortbl}
\usepackage[normalem]{ulem}

\begin{document}

\title{Chase Anonymisation: Privacy-Preserving Knowledge Graphs with Logical Reasoning
\thanks{The work of Pierangela Samarati was supported in part by the EC under project GLACIATION (101070141) and by project SERICS (PE00000014) under the MUR NRRP funded by the EU - NGEU. 
The views and opinions expressed in this paper are those of the authors and do not necessarily reflect the official policy or position of Banca d'Italia, EU or the Italian MUR. 
}
}
\author{\IEEEauthorblockN{Luigi Bellomarini}
\IEEEauthorblockA{
\textit{Banca d'Italia}\\
\small{luigi.bellomarini@bancaditalia.it}}
\and
\IEEEauthorblockN{Costanza Catalano}
\IEEEauthorblockA{
\textit{Banca d'Italia}\\
\small{costanza.catalano@bancaditalia.it}}
\linebreakand
\IEEEauthorblockN{Andrea Coletta}
\IEEEauthorblockA{
\textit{Banca d'Italia}\\
\small{andrea.coletta@bancaditalia.it}}
\and
\IEEEauthorblockN{Michela Iezzi}
\IEEEauthorblockA{
\textit{Banca d'Italia}\\
\small{michela.iezzi@bancaditalia.it}}
\and
\IEEEauthorblockN{Pierangela Samarati}
\IEEEauthorblockA{
\textit{Universit\`a degli Studi di Milano}\\
\small{pierangela.samarati@unimi.it}}
}

\maketitle

\begin{abstract}
We propose a novel framework to enable Knowledge Graphs (KGs) sharing while ensuring that information that should remain private is not directly released nor indirectly exposed via derived knowledge, maintaining at the same time the embedded knowledge of the KGs to support business downstream tasks. Our approach produces a privacy-preserving KG as an augmentation of the input one via controlled addition of nodes and edges as well as re-labeling of nodes and perturbation of weights. We introduce a novel privacy measure for KGs, which considers derived knowledge, a new utility metric that captures the business semantics we want to preserve, and propose two novel anonymisation algorithms. 
Our extensive experimental evaluation, with both synthetic graphs and real-world datasets, confirms the effectiveness of our approach.
\end{abstract}


\section{Introduction}

\begin{figure}[t]
    \centering
  
    \includegraphics[width=0.7\linewidth]{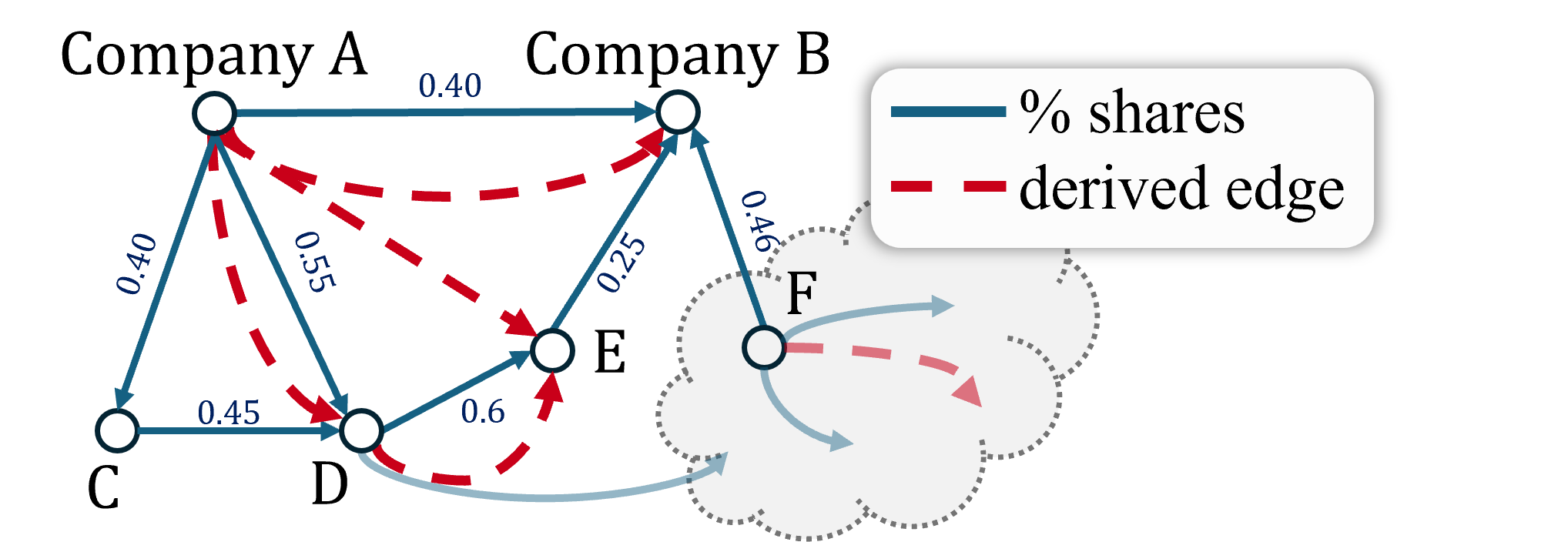}
    
    \caption{\small An ownership Knowledge Graph. 
    }\label{fig:knowledge_graph} 
\end{figure}

Knowledge Graphs (KGs) are gaining increasing scientific interest~\cite{DBLP:conf/i-semantics/EhrlingerW16,hogan2021knowledge}, as also witnessed by a broad adoption in many industrial domains, from healthcare to biotechnology, from logistics to finance~\cite{cao2022ai,magnanimi2023reactive,DBLP:journals/cacm/NoyGJNPT19,peng2023knowledge,potluru2023synthetic}.
While we are still lacking a consolidated and shared definition of KGs, a distinguishing characteristic is the presence of some form of intensional knowledge, an encoding of the business experience that can be leveraged to generate new---derived---nodes and edges, through a \textit{reasoning} process~\cite{bellomarini2019knowledge}. Logic-based approaches to KGs see nodes and edges as facts of a database (i.e., the \textit{extensional component}) and encode the business knowledge through a logic program (i.e., the \textit{intensional component}). The semantic of such a logic program is usually specified through the \textsc{chase} procedure~\cite{MaierMS79}, which applies the rules to the database, as long as they produce new facts (i.e., the \textit{derived extensional component}).
Consider for example Figure~\ref{fig:knowledge_graph}. It shows a company ownership KG: the nodes represent companies, and an edge from $x$ to $y$, with weight $w$, indicates that $x$ owns a fraction $w$ of the shares of $y$. The red edges are part of the derived extensional component, obtained by putting into action the business notion of \textit{company control}: \textit{a company $x$ controls $y$ if it owns more than $50\%$ of $y$'s shares, or if it controls a set of companies that jointly, and possibly with $x$ itself, own the majority of $y$.}

\noindent \textbf{Sharing Knowledge Graphs}.
Financial organisations, central banks, economic research entities, and national supervision authorities 
need to share KGs to address crucial business tasks, such as banking supervision, support to policymaking, anti-money laundering, and economic research. At the same time, the identity of the involved subjects and their relationships, be they individuals or companies, may not be disclosed~\cite{li2023private,majeed2020anonymization} while preserving the utility of the released data for downstream tasks~\cite{TADA_2025}.
This represents a severe hurdle, since classical approaches such as \textit{differential privacy} (DP)~\cite{chen2020publishing,jiang2021applications} and \textit{structural anonymisation} (e.g., k-anonymity~\cite{samarati2001protecting}) are unsuitable to handle KGs, where the intensional knowledge plays a crucial role and opens new avenues to privacy disclosure.

DP approaches are not applicable, as the injected noise undermines the effectiveness of downstream tasks, for example hampering the performance of machine learning models, especially those relying on multiple correlated graph statistics~\cite{hoang2023protecting}. 
Structural anonymisation techniques are more promising, but still insufficient for our goals. The methods of this family usually modify the topology of the graph, so that it exhibits at least $k$ ``similar'' (e.g., isomorphic) structures with respect to the adversary knowledge~\cite{hay2008,zhou2008}. 

Their typical setting is the so-called \textit{neighbourhood Attack Graph} (NAG), also known as a \textit{subgraph-based attack}~\cite{zhou2008}: the attacker tries to use some knowledge about the neighbourhood of a target node to re-map it in the original graph and re-identify it. The graph is modified to create, for each substructure of a given size, many undistinguishable isomorphic clones. Unfortunately, these techniques fail when the attacker has information on some derived knowledge and, in the worst case, they have full visibility of the business rules.

More in general, anonymising KGs presents various unique challenges, originating from the presence of derived knowledge, which we will discuss in detail: \emph{(i)} if the attacker knows the logical derivation steps (chase graph), even isomorphic structures in the graph can be distinguished, since, intuitively, their elements have a different ``lineage''; as a consequence, anonymising a KG does not trivially consist of pre-materializing the derived knowledge and adopting existing techniques on the obtained graph \emph{(ii)} in preserving the ``meaningfulness'' of the KG, the anonymisation process should account for the accuracy of business tasks 
\emph{(iii)} financial applications have high-levels privacy requirements, which cover not only the names of the entities, but also their ``role'' and relationships in the network.\\

In this work, we propose a \textit{reasoning-aware anonymisation approach} for Knowledge Graphs, where the derived knowledge is a first-class citizen. In particular, we introduce the \emph{$(k,x)$-chase anonymisation}, a novel structural anonymisation technique producing KGs that are NAR-resistant (resistant to NAG + Reasoning). Our technique modifies the KG so that each induced subgraph of size at most $x$ has other $k$ \emph{chase-isomorphic} structures, i.e. structures that are also indistinguishable with respect to the knowledge produced by the reasoning process (challenge~\emph{(i)}).
Our structurally indistinguishable subgraphs exhibit different node labels, in-degrees, and out-degrees, along with perturbed edge weights (challenge~\emph{(iii)}). In detail, this paper contributes the following.

\begin{itemize}[leftmargin=1em]
    \item The definition of a \textbf{$(k,x)$-chase anonymisation} for a KG, resistant to attacks where there is information of the reasoning rules or derived knowledge.
    \item A novel \textbf{semantic utility metric}, which accounts for the usefulness of the anonymised KG for specific downstream tasks (challenge~\emph{(ii)}). 
    Our anonymisation procedure maximises this metric and guarantees that the anonymised KG resembles the original one for the tasks of interest.  
    \item Two new \textbf{anonymisation algorithms}, namely, \emph{{\klone}} and \emph{{\kguard}}, able to reach a $(k,x)$-chase anonymisation while maximising the semantic utility metrics. The former is our first technique, which clones and differentiates the graph structures $k$ times, regardless of the size $x$ of the attack. The latter is a second approach that minimises the number of modifications by exploiting already existing structures.
    \item An \textbf{extensive experimental evaluation} on well-known network models and various real-world datasets, including the ownership KG of the Italian Central Bank~\cite{magnanimi2023reactive}.
\end{itemize}

\noindent \textbf{Overview}.

Section~\ref{sec:sketch}: problem definition and sketch of our approach. Section~\ref{sec:background}: background. Section~\ref{sec:pb_form}: adversary attacks. Section~\ref{sec:kx_iso}: definition of chase anonymisation. Sections~\ref{sec:baseline} and~\ref{sec:kguard}: anonymisation algorithms. Section~\ref{sec:experiments}: experimental evaluation. Section~\ref{sec:relwork}: related work. Section~\ref{sec:conclusion}: conclusions.

\section{Problem definition and sketch of the approach}\label{sec:sketch}
We consider the problem of releasing a privacy-preserving version of a Knowledge Graph protecting the actual identities of the entities (nodes)
in the graph and the relationships between them, even against attackers that have knowledge of portions of the original KG and reasoning rules on it or some derived knowledge. At the same 
time we aim at maintaining the utility of the privacy-preserving KG for downstream tasks for which it is released. 

\begin{figure*}
    \centering

 \begin{subfigure}{0.167\textwidth}
     \centering
    \includegraphics[width=0.6\linewidth]{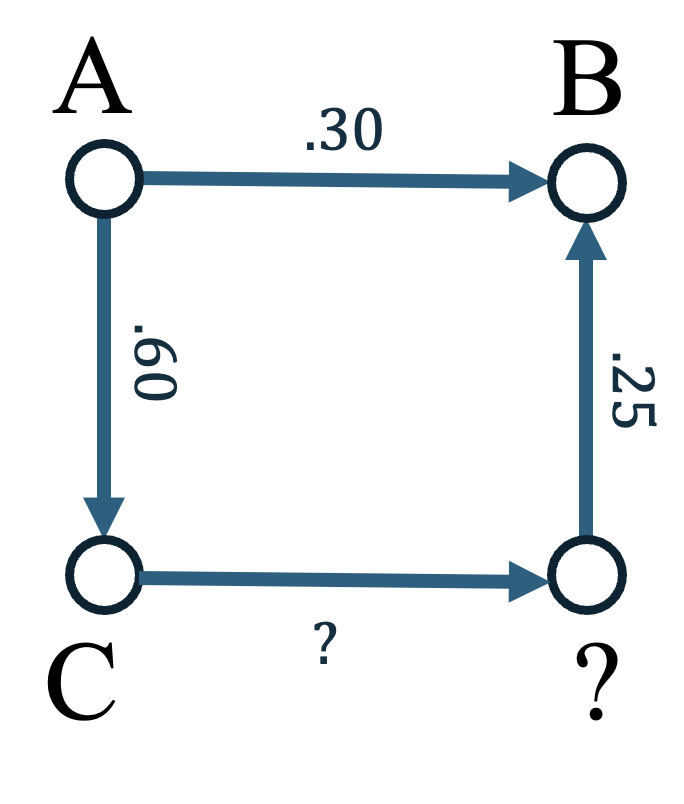}
    \caption{NAG (subgraph).}\label{fig:ex1_NAG}
    \end{subfigure}
     \hfill
    \begin{subfigure}{0.33\textwidth}
    \includegraphics[width=0.9\linewidth]{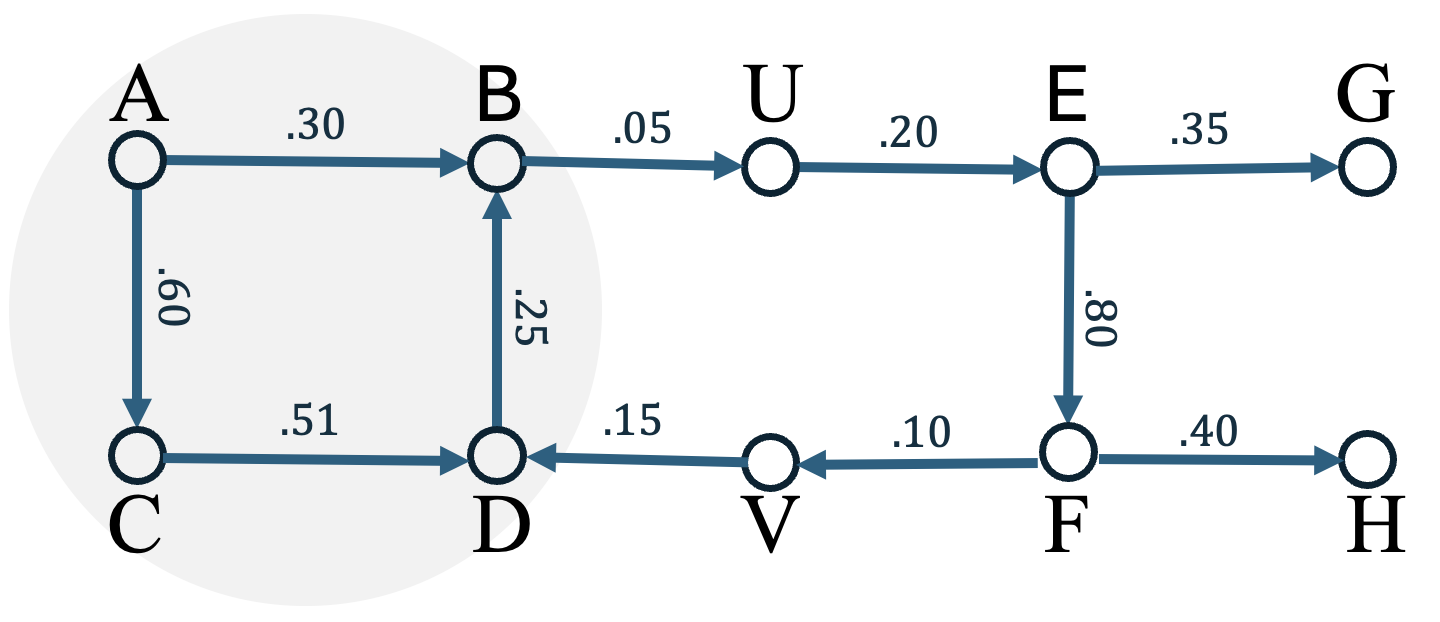}
    \caption{Real Knowledge Graph.}\label{fig:ex1_real_KG_a}
    \end{subfigure}
        \hfill
     \begin{subfigure}{0.33\textwidth}
      \centering
    \includegraphics[width=0.9\linewidth]{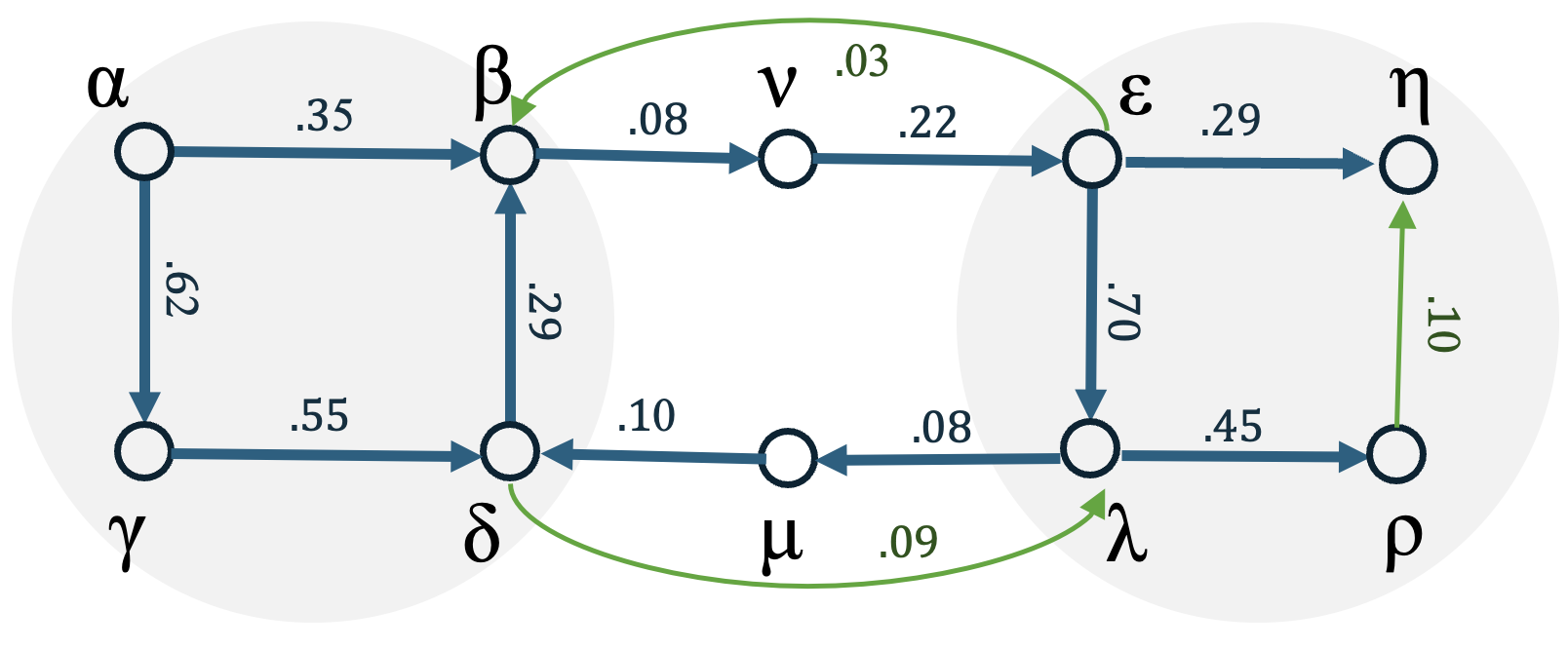}
    \caption{Structural anonymity.}
    \label{fig:ex1_kanonym}     \end{subfigure}

    \caption{\small 
    (a) Attacker's knowledge: an employee of C is aware that company A owns $60\%$ of their company's shares and $30\%$ of the shares of a third company B; they also know that C is acquiring the majority of the shares of an unknown company (?), which in turn holds $25\%$ of shares of B. (b) The NAG matches a uniquely identifiable structure within the real KG (grey shaded circle), revealing sensitive information. (c) Traditional structural anonymity prevents re-identification by introducing isomorphic structures, using additional edges (green edges).
    } 
    \label{fig:attacks}
\end{figure*}

As running example, we consider the KG in Figure~\ref{fig:ex1_real_KG_a}, where edges between nodes represent company ownership,
labels on the edges represent the share of such ownership.
We consider the set of rules formalised in Vadalog \cite{bellomarini2018vadalog} in Table~\ref{tab:rules}. A 
first set of rules defines company control, namely: 
a company $x$ controls itself ($\sigma_1$) and any other company $z$ for which it collectively has
more than 50\% either directly or through any company it controls, summing up all the edge share ($\sigma_2$).
A second set of rules defines reachability, meaning direct ($\sigma_3$) or indirect ($\sigma_4$) existence of a path from one node 
to the other; we will elaborate  on queries when discussing utility and experimental results. 
Classical structural anonymity would protect the KG by aliasing the identity of the nodes, perturbing edge labels, and 
possibly adding synthetic edges to ensure, for any subgraphs of KG of up to a given size $x$ (adversary's strength), the existence of at least 
$k$ (where $k$ is the privacy guarantee) isomorphic subgraphs. 
Structural anonymity protects against attackers knowing (from external knowledge) portions of the original graph and aiming 
at remapping it in the released KG to discover additional information. 
For instance, the privacy-preserving KG in Figure~\ref{fig:ex1_kanonym} protects against an attacker knowing Neighbourhood
Attack Graphs (NAGs) comprising up to four nodes with $k=2$. Indeed, the NAG in Figure~\ref{fig:ex1_NAG} has at least 2 isomorphic subgraphs (greyed) in such KG.

\begin{figure*}
    \centering

 \begin{subfigure}{0.13\textwidth}
     \centering
    \includegraphics[width=0.8\linewidth]{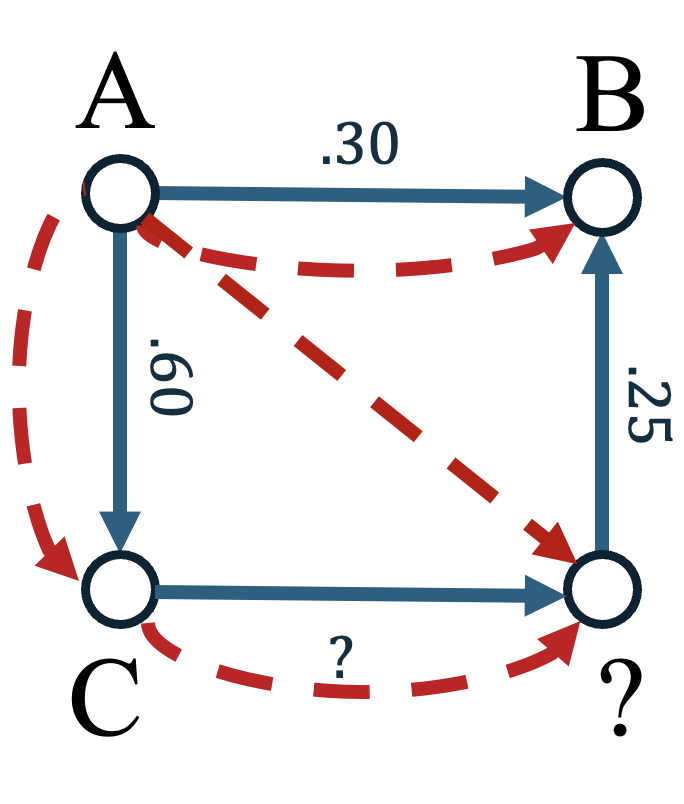}
    \caption{NAR.}\label{fig:ex1_NAR}
    \end{subfigure}
     \hfill
    \begin{subfigure}{0.28\textwidth}
    \includegraphics[width=1.1\linewidth]{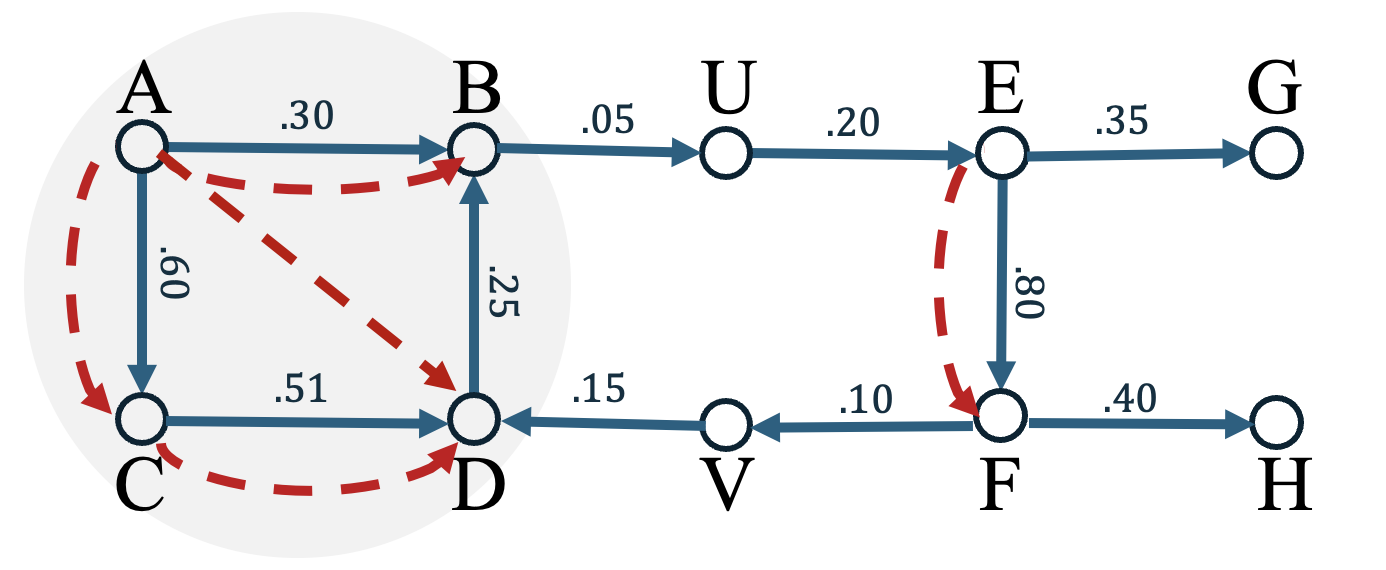}
    \caption{Real KG with derived edges.}\label{fig:ex1_real_KG_w_derived}
    \end{subfigure}
    \hfill
  \begin{subfigure}{0.28\textwidth}
  \centering
    \includegraphics[width=1.1\linewidth]{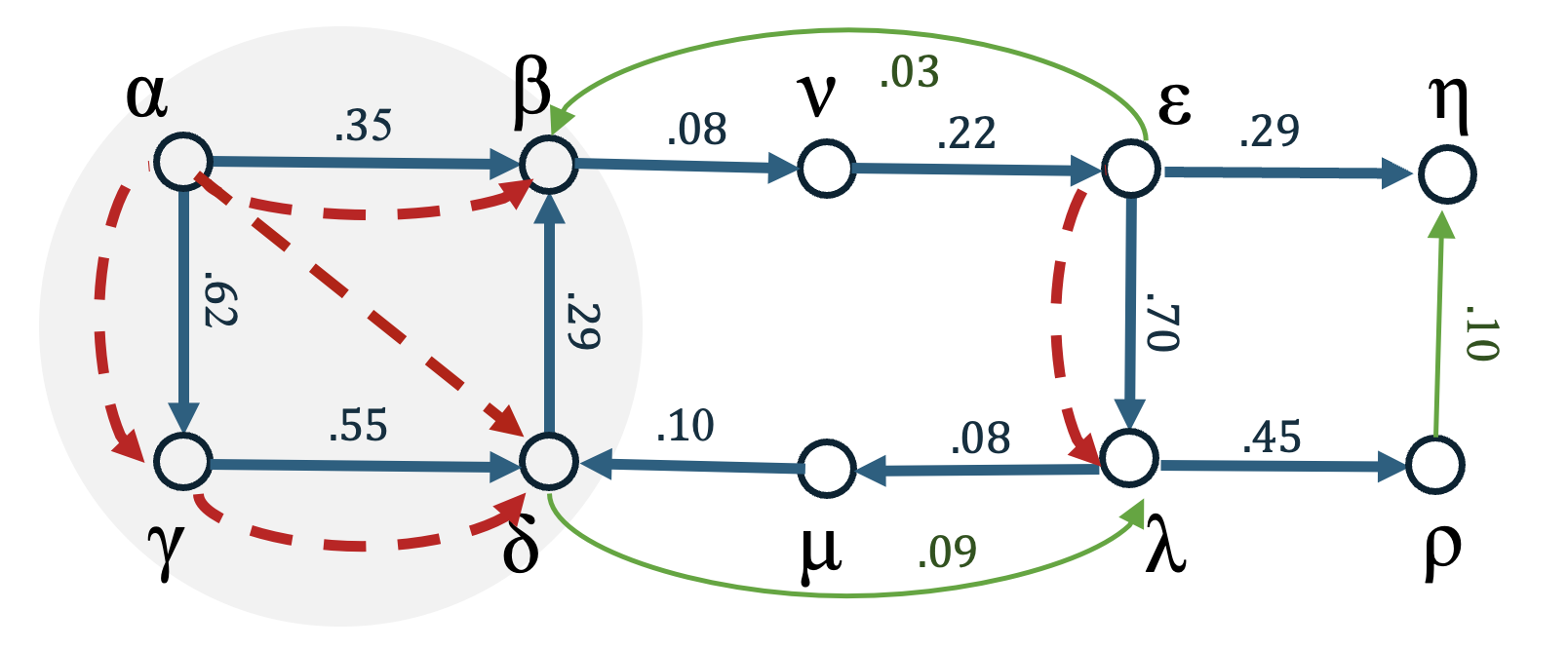}
    \caption{Failure of structural anonymity.}\label{fig:ex1_reidentification}
    \end{subfigure}
    \hfill
     \begin{subfigure}{0.28\textwidth}
      \centering
    \includegraphics[width=1.1\linewidth]{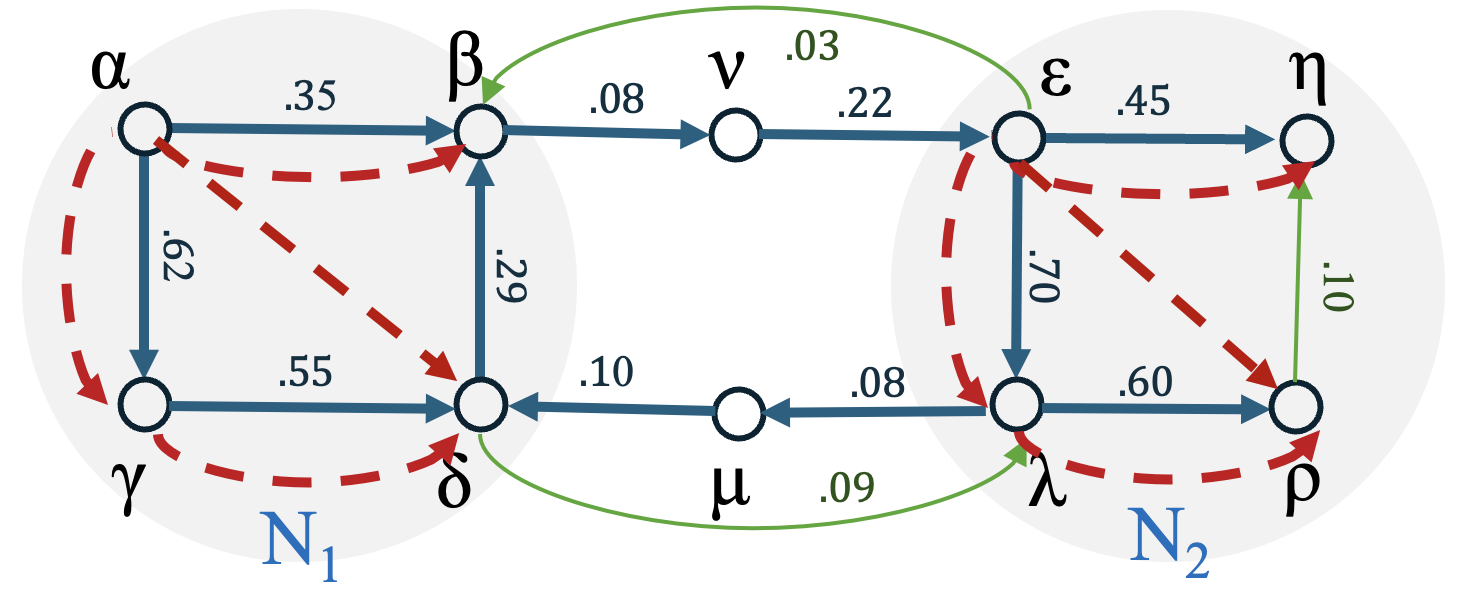}
    \caption{Chase anonymisation.} \label{fig:ex1_chase_iso_4a}
    \end{subfigure}
    \caption{\small Anonymisation of KGs with reasoning. (a) The employee of C 
    has full knowledge of the control rules, maybe grasped from a thoughtful reading of regulations, discovering all the control relationships and underlying facts that generate them. (b) The NAR matches a uniquely identifiable structure within the real KG. (c) Traditional structural anonymity fails, as the use of reasoning reveals the unique structure of the NAR within the graph (grey shaded circle). (d) Our \emph{chase anonymisation} addresses this limitation by explicitly considering the reasoning into the anonymisation procedure. 
    } 
    \label{fig:attacks}
\end{figure*}
\begin{figure*}
    \centering
\begin{subfigure}{0.3\textwidth}
      \centering
    \includegraphics[width=\linewidth]{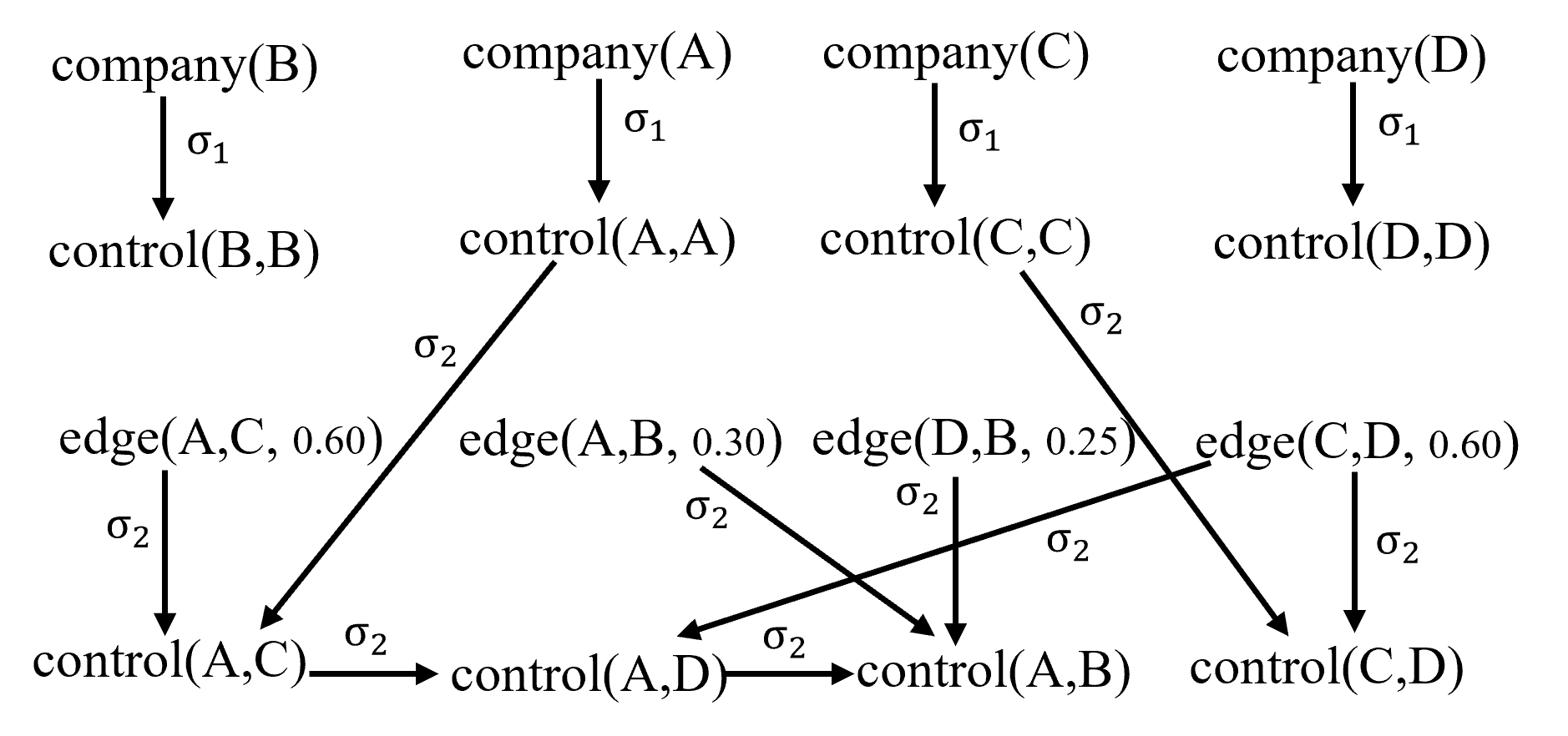}
    \caption{$V=\{A,B,C,D\}$.} \label{fig:ex1_chase_NAR}
    \end{subfigure}
    \begin{subfigure}{0.3\textwidth}
      \centering
    \includegraphics[width=\linewidth]{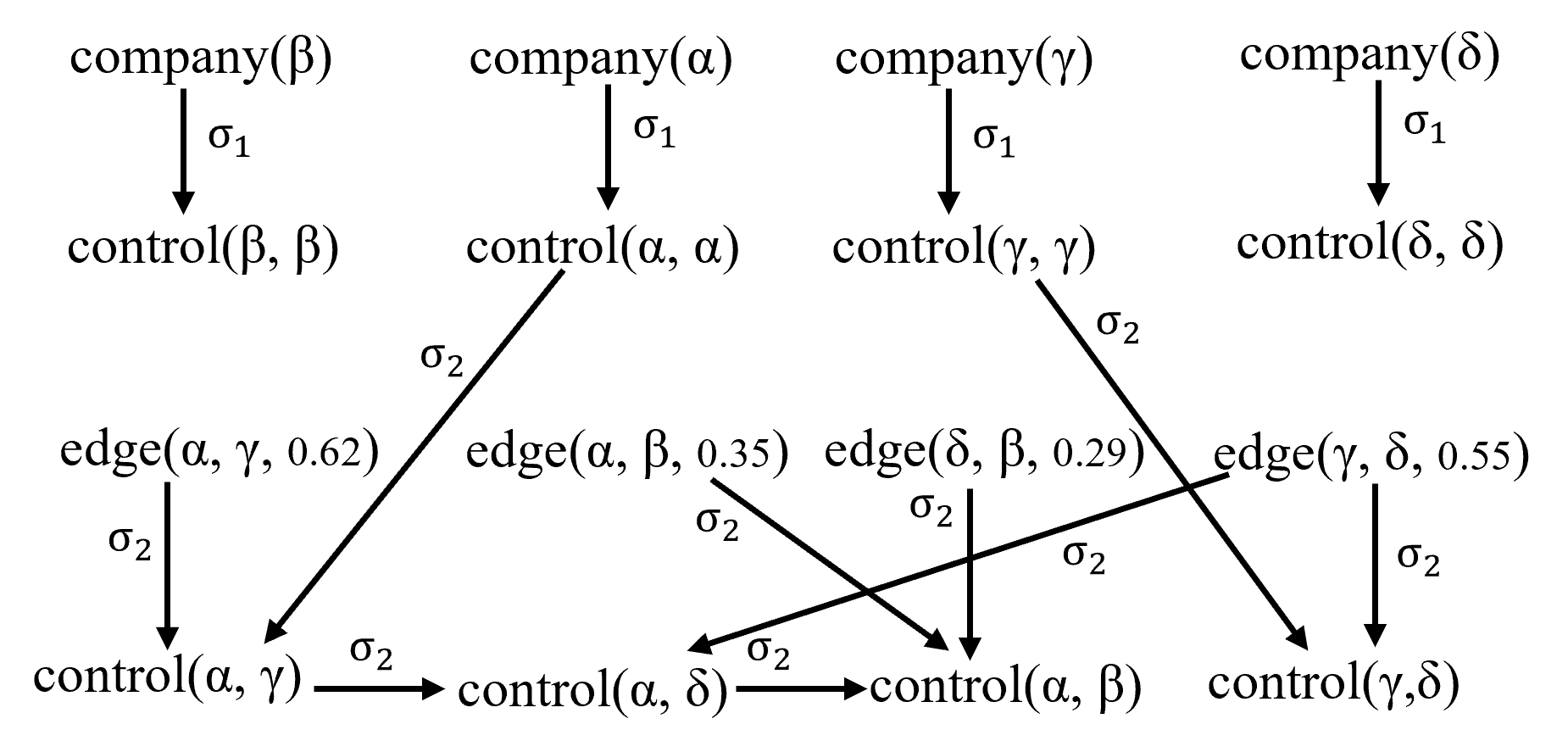}
    \caption{$V=\{\alpha, \beta, \gamma, \delta\}$.}
    \label{fig:ex1_chase1}
\end{subfigure}
     \begin{subfigure}{0.3\textwidth}
      \centering
    \includegraphics[width=\linewidth]{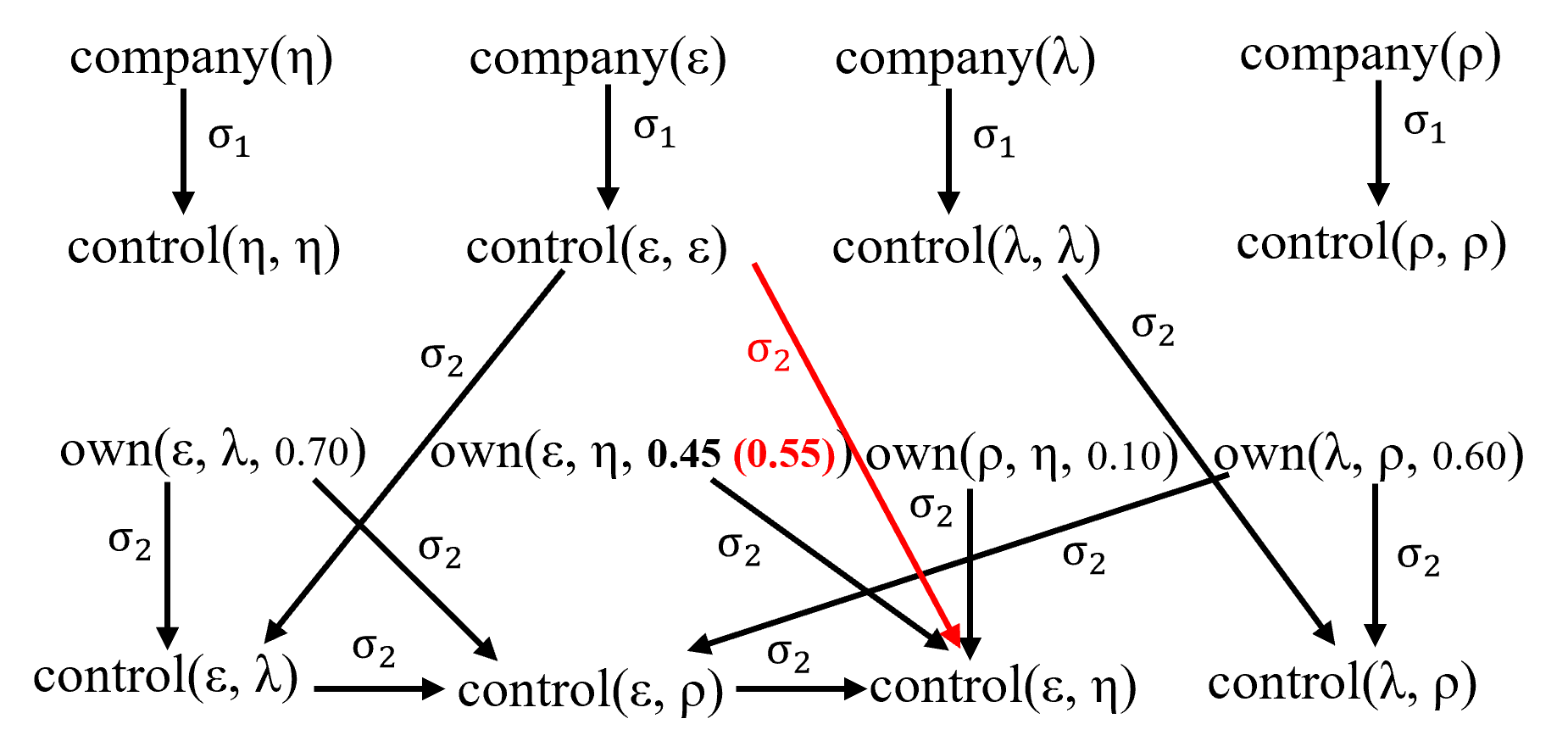}
    \caption{$V=\{\varepsilon, \eta, \lambda, \rho\}$.} \label{fig:ex1_chase2}
    \end{subfigure}
    \caption{\small (a) Chase graph of the NAR of Figure \ref{fig:ex1_NAR}; question marks are substituted by a placeholder "D" for the node and by random number $>0.5$ for the edge weight. (b) Chase graph of the subgraph $N_1$ of Figure \ref{fig:ex1_chase_iso_4a} (c) Chase graph of the subgraph $N_2$ of Figure \ref{fig:ex1_chase_iso_4a} (black edges). The red edge indicates how the chase graph changes if the weight of $(\varepsilon,\eta)$ is set to $0.55$. Subgraphs are induced by the vertex set $V$.}
     \label{fig:ex1_chase}
\end{figure*}
However, structural anonymity falls short when the attacker has information on the reasoning rules producing derived 
knowledge from KG. 
In fact, knowledge of the  rules and hence of the derived edge allows the attacker to 
tell apart isomorphic subgraphs. This is illustrated in Figure~\ref{fig:ex1_NAR} where the knowledge derived from the
application of the rules (red dotted edges) enables distinguishing between the two subgraphs that were structural isomorphic in Figure~\ref{fig:ex1_kanonym}, thus enabling the exact remapping of the NAR in the sanitized KG. 

Our chase anonymisation aims at guaranteeing privacy protection against such NAR attacks, in which attackers have knowledge not only
of a subgraph of the ground extensional component of the KG (NAG), but also of the set of reasoning rules.
Ensuring such protection requires operating on the KG so to ensure isomorphism not only with respect to the structure but also
with respect to the additional edges (derived knowledge). Even more, it requires ensuring indistinguishability with respect to
the sequences of facts (new knowledge) produced by the application of the rules to the graph, i.e., with respect to their \emph{chase graphs}.

Given a KG and a set of rules, the chase graph is a graph whose nodes represent 
the (intensional or derived) facts and edges connects facts to other facts which they contributed to derive;
a label on the edge reports the rules enabling such derivation. 
For instance, Figure~\ref{fig:ex1_chase1} reports the chase graphs of the left grey subgraph in Figure~\ref{fig:ex1_reidentification}.
The anonymized graph in Figure~\ref{fig:ex1_chase_iso_4a} satisfies the chase anonymisation as the two grey subgraphs $N_1$ and $N_2$ in it, 
not only have the same derived knowledge, but also the same (apart from renaming) chase graphs (Figures \ref{fig:ex1_chase1} and \ref{fig:ex1_chase2}, only black edges).
However, had the weight between $\varepsilon$ and $\eta$ in $N_2$ been set to $0.55$ instead of $0.45$, 
while the two subgraphs remain isomorphic, their chase graphs would have differed,
the one of $N_2$ having one additional edge (red edge in Figure~\ref{fig:ex1_chase2}).

In our approach we develop two algorithms that, for each induced subgraphs of size $x$, guarantee the presence of other $k$ chase-isomorphic subgraphs within the anonymised KG. Moreover, labels and in- and out-degrees of the vertices are different between isomorphic copies, ensuring the privacy of entities and their relations. Also edge weights are perturbed with respect to the original KG preventing their use for re-identification. Lastly, we optimise the anonymisation process (in the weight perturbation and weight assignment to the newly introduced synthetic edges) by defining a set of business tasks on which we want the anonymised KG to be as close as possible to the original one. The business tasks are modeled as conjunctive queries (as the ones in Table \ref{tab:rules}, for its semantics see Section \ref{sec:background}), and a Jaccard-based similarity index is set up to measure the difference between the answers returned by the anonymised and original KG, which is dynamically optimised in the proposed algorithms. For its formal definition and results
we refer the reader, respectively, to Section \ref{sec:kx_iso} and Section \ref{sec:experiments}.\\

We also consider a lighter attack where the attacker has information on some derived edges but not the set of reasoning rules. In this case, we show that the anonymisation algorithms can be simplified to make the anonymised graph resistant to these types of attacks (but vulnerable to NAR).

\section{Basic concepts}\label{sec:background}
\textbf{Notation and graph concepts.}

We set $[n]:=\{1,2,\dots, n\}$. 
The restriction of a function $f:B\to C$ to a subset $A\subseteq B$ is denoted by $f_{|_A}$; the identity function on a set $A$ is indicated by $id_A$.
A (labelled weighted directed) graph is a tuple $G=(V,E,L,\omega, l)$ where $V$ is the set of vertices, $E\subseteq V\times V$ is the set of edges, $L$ is the set of labels, $\omega:E\to [0,1] $ is the edge weight function and $l:V\to L $ is the vertex label function.
Selfloops, i.e. edges of type $(v,v)$, are allowed.
The cardinality $|V|$ of the vertex set is indicated by $n$.
 Given $v \in V$, we set 
 $N^{-}\sub G(v):=
     \{u\in V:(u,v)\in E\}$ and 
 $N^{+}\sub G(v):=
     \{u\in V: (v,u)\in E \}$. 
 \normalsize
 The in-/out-degree of a vertex $v$ are defined, respectively, as $d^{-}\sub G(v):=|N\sub G^{-}(v)| $ and $d^{+}\sub G(v):= | N\sub G^{+}(v)|$. When clear from the context, we may omit the subscript $G$.
 A directed graph is \textit{weakly connected} if, by replacing all of its directed edges with undirected ones, there exists a path connecting any pair of vertices~\cite{bang2008}.
  We write $V(G)$, $E(G)$, $L(G)$, $\omega\sub G$ and $l\sub G$ to denote, respectively, the vertex set, the edge set, the label set, the weight function, and the label function of a graph $G$.
Given $X\subseteq V $, the subgraph of $G$ \emph{induced} by $X$ is given by $G[X]:=(X,E\sub X,L, \omega_{|\sub {E{\sub X}}},l _{|_{X}})$, where $E\sub X=\{(u,v)\in E: u,v\in X \}$. 
Two graphs $G$ and $G'$ are \emph{isomorphic} through the \emph{isomorphism} $\psi:V(G)\to V(G')$ if $\psi$ is bijective and $(u,v)\in E(G)$ iff $(\psi(u),\psi(v))\in E(G') $.\\
\textbf{Relational foundations and Knowledge Graphs.}
Let $\mathbf{C}$, $\mathbf{N}$, and $\mathbf{V}$ be disjoint countably infinite sets of {\em constants}, {\em (labelled) nulls} and {\em variables}, respectively.  
They are also known as {\em terms}. 
A {\em (relational) schema} $\mathbf{S}$ is a finite set of predicates with associated arities. An  {\em atom} is an expression $R(\bar v)$, where $R\! \in\! \mathbf{S}$ is of arity $n \!\geq\! 0$ and $\bar v$ is an $n$-tuple of terms. A {\em database} $D$ over $\mathbf{S}$ associates to each relation symbol in $\mathbf{S}$ a relation of the respective arity over the domain of constants and nulls. We denote by $\textit{dom}(D)$ the set of constants in $D$.
Relation members are called \textit{tuples} or \textit{facts}. 
Datalog$^\pm$ rules~\cite{CaGL09} are first-order implications $\forall \mathbf{x}~\boldsymbol{\phi}(\mathbf{x}) \to \exists\mathbf{z}~\boldsymbol{\psi}(\mathbf{y},\mathbf{z})$, where $\boldsymbol{\phi}(\mathbf{x})$ (the \textit{body}) and $\boldsymbol{\psi}(\mathbf{y},\mathbf{z})$ (the \textit{head}) are conjunctions of atoms over $\mathbf{S}$ and boldface variables denote vectors of variables, with $\mathbf{y}\subseteq\mathbf{x}$. We write these existential rules as $\boldsymbol{\phi}(\mathbf{x}) \to \exists\mathbf{z}~\boldsymbol{\psi}(\mathbf{y},\mathbf{z})$, using commas to denote conjunction of atoms in $\boldsymbol{\phi}(\mathbf{x})$ and $\boldsymbol{\psi}(\mathbf{y},\mathbf{z})$. 
The semantics of a set of Datalog$^\pm$ rules $\Sigma$ applied to a database $D$, can be defined operationally through the chase procedure~\cite{MaierMS79}. The chase iteratively expands the database $D$ with new facts, possibly containing labeled nulls, into a database $ chase(D,\Sigma)$ by iteratively
applying the rules $\Sigma$ to it until a fixed point is reached. 
We adopt the Vadalog~\cite{bellomarini2018vadalog} syntax, a restriction of Datalog$^\pm$ that incorporates all the mentioned features while assuring convergence of the process. The \emph{chase graph} $\mc G(D,\Sigma)$ is a directed graph having
nodes labelled after facts from $chase(D,\Sigma)$ and having an edge from a node $a$ to a node $b$ labeled by $\sigma$ if the fact $b$ is obtained from $a$ and possibly from other facts by the application of rule $\sigma\!\in\! \Sigma$ (a chase
step). 
We consider databases that can be represented in terms of a weighted labelled directed graph $G$. A Knowledge Graph (KG) is a pair $(G,\Sigma)$ with $G$ a weighted labelled directed graph (\textit{ground extensional component}) and $\Sigma $ is a finite set of Vadalog rules (\textit{intensional component}).\footnote{In this paper we focus on KGs having edge weights in [0,1] as our industrial
application is of this type. We note however that our approach can be extended to a
more general case of labelled edges, assuming the specification of an ontology capturing
labels and distance between them to enable label perturbation while maintaining utility.
Specifically, Algorithms \ref{alg:baseline} (Section \ref{sec:baseline}) and \ref{alg:algo_main} (Section \ref{sec:kguard}) can be adapted
 to this general case by substituting the function WeightNoising with any other function that perturbs the edge’s labels of the KG while maintaining high utility.
} 
The new facts produced by the application of $\Sigma$ on $G$ are mapped into new edges for $G$ (\textit{derived extensional component}), also called \emph{derived} edges. The set of derived edges is denoted by $\mathcal{D}_{_{G,\Sigma}}$ and no weight is assigned to them. 
In Figure \ref{fig:knowledge_graph}, $ \mathcal{D}_{_{G,\Sigma}}$ is the set of red edges, derived from the application of the control rules $\Sigma=\{\sigma_1,\sigma_2\}$ of Table \ref{tab:rules}. 
We require $\Sigma$ to be such that the facts produced on a smaller instance of a database $D$ to be facts also produced when performed on the entire $D$. This is the case, e.g., of the control and reachability rules in Table \ref{tab:rules} but does not necessarily hold true. Formally, 
given a KG $(G,\Sigma)$, $\Sigma$ is said to be \emph{monotonic} on $ G $ if for all $X\subseteq Y\subseteq V $, we have that 
$\mc G(G[X],\Sigma)\subseteq  \mc G(G[Y],\Sigma) $.
Without loss of generality, in this paper we consider KGs with a weakly connected ground extensional component;
general cases can be dealt with by applying our method to each weakly connected component of the KG.\\
\textbf{Queries.} We express business tasks on KGs as queries, like the one in Table \ref{tab:rules}.
A \textit{conjunctive query} $Q$ over a schema $\mathbf{S}$ of a database $D$ is an implication $q(\mathbf{x}) \leftarrow \boldsymbol{\phi}(\mathbf{x},\mathbf{y})$, where $\boldsymbol{\phi}(\mathbf{x},\mathbf{y})$ is a conjunction of atoms over $\mathbf{S}$, $q$ is an n-ary predicate, and $\mathbf{x}$ and $\mathbf{y}$ are vectors of terms. In the presence of $\Sigma$, it is evaluated as a set of tuples as follows: $ Q(D)=\{t\in dom(D)^n: q(t)\in chase(D,\Sigma)\}$. 
\begin{table}
\caption{Reasoning rules and queries. The term $\text{edge}(x,y, w)$ represents an edge from entity $x$ to $y$ of weight $w$.}\label{tab:rules}
{\footnotesize
\begin{tabular}{l}
\toprule
\textbf{Reasoning rules: company control} \\

$\sigma_1: $  $\text{company}(x) \rightarrow \text{control}(x, x)$ \\
$\sigma_2: $  $\text{control}(x, y), \text{edge}(y, z, w), v = \text{sum}(w), v > 0.5\rightarrow  \text{control}(x, z)$ \\
\midrule
 \textbf{Reasoning rules: reachability} \\

$\sigma_3: $  $\text{edge}(x, y, w ), x \neq y, w>0\rightarrow \text{reach}(x, y)$ \\
$\sigma_4: $  $\text{reach}(x, z), \text{edge}(z, y, w), x \neq y \neq z, w>0 \rightarrow \text{reach}(x, y)$ \\
\midrule
\textbf{Queries}   \\

$Q_1 :$  $\text{reach-03}(x,y) \leftarrow \text{edge}(x, y, w), x \neq y, w \geq 0.3 $\\
$Q_2 :$  $\text{reach}_\rightleftarrows(x,y) \leftarrow \text{edge}(x, y, w), \text{edge}(y, x, o),  x \neq y, w > o$ \\

  $Q_3 :$  $2\text{-edge}(x)\leftarrow  \text{edge}(x, y, w), x \neq y, k = \text{sum}(1, \langle y \rangle), k \geq 2$ \\
$Q_4 :$  $2q\text{-edge}(x)\leftarrow \text{edge}(x, y, w), x \neq y, k = \text{sum}(1, \langle y \rangle), w > q, k \geq 2$ \\
$Q_5 :$  $\text{direct-control}(x,y)  \leftarrow \text{edge}(x, y, w), x \neq y, w > 0.5$ \\
$Q_6: $  $\text{chain-control}(x,y)\leftarrow \text{control}(x, y), \text{edge}(x,t,\_), \text{edge}(t,z, \_),$\\
  $\qquad\qquad\qquad\qquad\qquad\qquad\qquad\qquad\quad\text{edge}(z,y, \_), x \neq  y\neq t\neq z.$ \\
$Q_7 :$  $\text{ultimate-control}(x,y)  \leftarrow \text{control}(x,y), \neg \text{control}(\_,x) $ \\

\bottomrule
\end{tabular}
}
\end{table}

\section{Adversary attacks}\label{sec:pb_form}

Our adversary model extends the classical assumption of adversaries knowing a portion of the knowledge graph, called Neighbourhood Attack Graph (NAG), by providing also protection against adversaries knowing the reasoning rules or some derived knowledge.
We consider two kinds of adversary attacks. The first attack, denoted as NAR (for NAG + Reasoning)
models the worst case scenario in which the attacker knows a NAG as well as the Reasoning rules. 
The second attack, denoted as NAD (for NAG + Derived), models the case where the attacker knows a NAG 
and some derived edges.

\textbf{NAR attack.}
The first scenario we consider is where the adversary knows: 
\begin{enumerate}
    \item the topology of a weakly connected subgraph of the KG;
   \item the label of at least one vertex of the subgraph;
   \item the weight of at least one edge of the subgraph;
   \item the set $\Sigma$ of reasoning rules.
\end{enumerate}

Given $X$ a set of $x$ vertices, the maximal knowledge that an attacker may have in terms of items (1)--(4) is to have information on the full topology of the induced subgraph $G[X]$, together with all its vertex labels and edge weights, of the set $\Sigma$ of reasoning rules, which implies also the knowledge of all the derived edges of the subgraph. We are going to formalise the attack in this worst-case scenario.

\begin{definition}[NAR]\label{def:NAG}
Given $(G,\Sigma)$ a KG,
the information possessed by the adversary is a pair $M=(G[X], \Sigma) $, where  $X$ is a subset of vertices of $G$ and $G[X]$ is weakly connected. We call $M$ the \emph{neighbourhood Attack graph with Reasoning (NAR)} of size $x=|X|$.

\end{definition}

The knowledge of the NAR $ (G[X],\Sigma)$ implies the knowledge of the chase graph $\mc G(G[X],\Sigma)$.

\textbf{NAD attack.}

We consider a second weaker scenario in which the adversary knows items (1)--(3) as in the previous NAR attack model, but instead of having information on the full set of reasoning rules $\Sigma$, they have knowledge of 
\begin{enumerate}
   \item[(4')] some derived edges between the vertices of the subgraph. 
\end{enumerate}

In this scenario, the adversary has knowledge only of the (possibly partial) results of the application of the rules of $\Sigma$ to the subgraph, i.e. the derived edges.  
Given $X$ a set of $x$ vertices, the maximal knowledge that an attacker may have in terms of items (1)--(4') is to have information on the full topology of the induced subgraph $G[X]$, together with all its vertex labels, edge weights and derived edges $D\sub{{G[X],\Sigma}}$. We formalise the attack in this worst-case scenario. 

\begin{definition}[NAD]\label{def:NAG}
Given $(G,\Sigma)$ a KG,
the information possessed by the adversary is a pair $N=(G[X], \mc D\sub{{G[X],\Sigma}} ) $, where $X$ is a subset of vertices of $G$ and $G[X]$ is weakly connected. We call $N$ the \emph{neighbourhood Attack graph with Derived edges (NAD)} of size $x=|X|$.

\end{definition}

We protect the KG from a NAR (and hence NAD) attack of size $x$, by guaranteeing that in the anonymised KG each subgraph of size at most $x$ has at least other $k-1$ subgraphs that are chase-isomorphic, i.e. that are indistinguishable not only in their topology (including derived edges), but also in the sequence of facts generated by the reasoning. In this way, the adversary cannot re-identify the structure, as it has no information on how to discern between them. 
Moreover, we guarantee that the corresponding vertices in the isomorphism have all different labels, in-degree and our-degree, thus ensuring that such information cannot be deduced by the adversary. Like in structural anonymity, our protection degree $k$ models the privacy guarantee (uncertainty of re-mapping) against the adversary.

\section{\((k,x)\)-chase anonymisation}\label{sec:kx_iso}

In this section, we illustrate our $(k,x)$-chase anonymisation 
to protect against the attacks previously described. We first model protection 
against the most powerful NAR attack and then show how it can 
be simplified if only NAD attacks are to be considered.

\textbf{Protection from NAR attacks.}

Our approach to protect against NAR attacks relies on the following new concept of chase-isomorphism.

\begin{definition}[Chase-isomorphism]\label{def:chase_isomorphism}
Let $(G,\Sigma)$ and $(G',\Sigma)$ be two KGs with the same set of reasoning rules and vertex sets $V$ and $V'$ respectively.  
We say that they are \emph{chase-isomorphic} ($G{\cong}\sub{\Sigma} G'$) if:
\begin{enumerate}
\item  their chase graphs $\mc G(G,\Sigma )$ and $\mc G(G',\Sigma )$ are isomorphic through the isomorphism $\Psi$;
\item $\Psi_{|_{V}}$ is a bijection between $V$ and $V'$; 
\item for each edge $(a,b)$ of $\mc G(G,\Sigma ) $, $\sigma$ is the label of $(a,b)$ if and only if $\sigma$ is the label of $(\Psi(a),\Psi(b))$ in $\mc G(G',\Sigma ) $.
\end{enumerate}
\end{definition}
I.e.\ , two KGs are chase-isomorphic if at each chase step they produce the same facts, up to a re-labeling of the vertices.

Clearly, KG anonymisation requires anonymising the nodes in the KG. In this respect we note that what makes nodes identifiable 
is not only their label (which can be aliased) but also their topological characteristics in the graph (i.e., in- and out-degree) that can 
make a node recognisable. We denote the triple comprising label and in- and out-degree as the sensitive attributes of a node, 
as captured by the following definition.

\begin{definition}[{Sensitive Attributes}]\label{def:privacy}
Given a KG $(G,\Sigma)$ and a node $v\in V$, the \emph{sensitive attributes} of $v$ is the triple:

$\xi\sub G(v) = \{ l\sub G(v), d\sub G^{-}(v), d\sub G^{+}(v)\}$, 
made of, respectively, its label, in-degree and out-degree. 
\end{definition} 

We write $\xi\sub G(u)\neq \xi\sub G(v)$ to indicate that $l\sub G(u)\neq l\sub G(v)$, $d\sub G^{-}(u)\neq d\sub G^{-}(v)$ and $d\sub G^{+}(u)\neq d\sub G^{+}(v)$. 
To protect the KG from a NAR attack, the idea is to require each induced subgraph of $x$ vertices to have at least other $ k-1$ chase-isomorphic subgraphs with different sensitive attributes. 
This guarantees the NAR to match at least with $k$ different subgraphs, each exhibiting different sensitive attributes: this implies that the attacker, if they were to choose a structure at random to match their NAR, would retrieve the correct structure with probability less than or equal to $1/k$. 
We call it a \emph{\((k,x)\)-chase anonymisation}.

More specifically, we want to anonymise the released graph by creating a new KG $(A,\Sigma)$, such that 1) $A$ is obtained by adding synthetic edges and vertices to $G$; 2) the vertex labels of $A$ do not coincide with the ones of $G$, so that the attacker cannot retrieve the real names of the entities; 3) the edge weights of $A$ are different from those of $G$, so that the attacker cannot use them for matching; 4) each induced subgraph of $G$ of size $x$ is (ii) chase-isomorphic to other $k-1$ induced subgraphs in $A$, which must be (i) pairwise vertex-disjoint and (iii) with different sensitive attributes. In particular, each node of a subgraph has different label, in-degree and out-degree with respect to the corresponding nodes in each of the $k$ isomorphic subgraphs.
This is formalised as follows.
\begin{definition}[(k,x)-chase anonymisation]\label{def:isomo}

Let $(G,\Sigma)$ be a KG with $G=(V,E,L,\omega, l)$ weakly connected and let $x\in [n]$ and $k\in \mathbb{N}$, $k\leq \binom{n}{x}$. A \emph{$(k,x)$-chase anonymisation} of $(G,\Sigma)$ is a KG $(A,\Sigma)$ with $A=(V\sub A,E\sub A,L\sub A,\omega\sub A, l\sub A)$ such that:
\begin{enumerate}
\item $V\subseteq V\sub A$, $E\subseteq E\sub A $; \emph{[augmentation]}

\item $L\cap L\sub A=\emptyset$; \emph{[anonymisation of labels]} 
\item $  \omega\sub A(e)\neq \omega(e) $ for all $e\in E $; \emph{[perturbation of weights]}

    \item for all $X_1\subset V$ such that $|X_1|=x$ and $G[X_1]$ is weakly connected, there exist $X_2, X_3, \ldots, X_{k} \subseteq V\sub A$ such that:
    \begin{enumerate}
        \item[(i)] $X_i\cap X_j=\emptyset $ for all $i,j\in [k]$, $i\neq j$; \emph{[vertex-disjointness]}
    \item[(ii)]  $A[X_1]\cong\sub \Sigma A[X_i]$ through the chase-isomorphism $\Psi_i$, for all $i\in [k]$ with $\Psi_1=id\sub {{X_1}}$; \emph{[anonymity]}
   \item[(iii)] $\xi\sub A(\Psi_i(v))\neq \xi\sub A(\Psi_j(v)) $ for all $v\in X_1$, $i,j\in [k]$ with $i\neq j$; \emph{[diversity]}
    \end{enumerate}
\end{enumerate}
\end{definition}

Here, parameter $x$ refers to the maximum size of the substructures we can resist the attack, while $k$ is related to the probability of correct re-identification (higher values guarantee stronger protection).
At the same time, high values of $k$ and $x$ might substantially increase the computational cost of the anonymisation (see Section \ref{sec:experiments}). 
Therefore, the choice of $k$ and $x$ is to be evaluated case by case, depending on the application and on the desired level of privacy. The following result guarantees that a $(k,x)$-chase anonymisation is also resistant to NAR attacks of smaller size; the proof can be found in the supplementary material \cite{appendix}.

\begin{proposition}\label{prop:smallerNAG}
Let $(A,\Sigma)$ be a $(k,x)$-chase anonymisation of a KG $(G,\Sigma)$ with $\Sigma $ monotonic. 
Then $(A,\Sigma)$ is a $(k,x')$-chase anonymisation of $(G,\Sigma)$ for any $x' \leq x$. 

\end{proposition}
\begin{proof}
Items 1)--3) of Definition \ref{def:isomo} are trivially fulfilled for the $(k,x')$-chase anonymisation. Let now consider $X'\subset V$ such that $|X'|=x'\leq x$ and $G[X']$ is weakly connected. Then there must exist $X_1\subseteq V$ such that $X'\subseteq X_1$, $|X_1|=x$, and $G[X_1]$ is weakly connected, since $G$ is weakly connected. By hypothesis, there exist $X_2, X_3, \ldots, X_{k} \subseteq V_A$ of cardinality $x$ that fulfil item 4). For each $i\in [k]$, let $ X'_i=\Psi_i(X')\subset X_i\subseteq V\sub A$; then $|X'_i|=x' $. We want to show that $X'_2,X'_3,\dots ,X'_k$ are the vertex sets that satisfy item (4) for the $(k,x')$-chase anonymisation. Item (4i) is satisfied since $X'_i\cap X'_j\subset X_i\cap X_j=\emptyset $ and also item (4iii) is trivially fulfilled. 
   Due to the monotonic property we have that 
$\mc G(G[X'],\Sigma) \subset \mc G(G[X_1],\Sigma)  $. Then the restriction of $\Psi_i$ to the nodes of $\mc G(G[X'],\Sigma) $ is an isomorphism between the chase graphs $ \mc G(G[X'],\Sigma)$ and $ \mc G(G[X'_i],\Sigma)$ with $\Psi_{|_{X'}}$ a bijective function between $X'$ and $X'_i$, thus satisfying item (4ii).
\end{proof}

Proposition~\ref{prop:smallerNAG} can be violated if $G$ is not weakly connected. In this case, a $(k,x)$-chase anonymisation $A$ does not guarantee that a weakly connected component $C$ of size $x'<x$ is chase-isomorphic to $k-1$ other subgraphs, as it is not contained in any weakly connected subgraphs of size $x$. This problem can be overcome by performing a $(k,x')$-chase anonymisation on every weakly connected component $C$ of $G$, with $x'=\min\{x,|C|\}$.
Conversely, the hypothesis of the NAR to be weakly connected could be removed. Indeed, suppose to have a $(k,x)$-chase anonymisation $A$ of $G$ and a NAR of size $x$ not weakly connected. Let $C_1,\dots ,C_m$ be its weakly connected components with $|C_j|=x_j$ for $j\in [m]$: then each $C_j$ is a weakly connected NAR with $x_j<x$. By Proposition~\ref{prop:smallerNAG}, $A$ is anonymised for any of such components, hence it is resistant to the original attack. Hence, a NAR that is not weakly connected can be equivalently seen as a sequence $C_1,\dots,C_m$ of weakly connected NARs of smaller size.

\textbf{Protection from NAD attacks.}
The $(k,x)$-chase anonymisation presented in the previous paragraph makes the KG resistant also to attacks of type NAD, as they involve less knowledge than NAR attacks. We here introduce a lighter anonymisation framework that is resistant only to a NAD attack, but which may result in fewer added structures. Thus, despite the worst case computational complexity remaining the same as in the NAR case, this anonymisation framework may lead to smaller computational costs in real-world scenarios (see Section \ref{sec:experiments}). It relies on the concept of \emph{KG-isomorphism}, extending the classical concept of graph-isomorphism.

\begin{definition}[KG-isomorphism]\label{def:isomorphism}
Let $(G,\Sigma)$ and $(G',\Sigma)$ be two KGs with vertex sets $V$ and $V'$ and edge sets $E$ and $E'$ respectively.
 
We say that they are \emph{KG-isomorphic} if there exists a bijective function $\Phi: V\to V'$, called \emph{KG-isomorphism}, such that for all $u,v\in V$:
\begin{align}
(u,v)\in E &\iff (\Phi(u),\Phi(v))\in E'\label{eq:iso1}\\
(u,v)\in \mc D_{_{G,\Sigma}} &\iff  (\Phi(u),\Phi(v))\in\mc D_{_{G',\Sigma}}\,\, .\label{eq:iso2}
\end{align}
\end{definition}

I.e.\ two KGs are KG-isomorphic if they share both the topology of the ground extensional component (eq.(\ref{eq:iso1}) refers to classic isomorphism between graphs) and on the derived component (eq.(\ref{eq:iso2}) enforces the correspondence also on the derived edges). 
If two  KGs are chase-isomorphic, then they are KG-isomorphic; the converse is not true, as shown in Fig.\ref{fig:ex1_chase}. 

If we are interested in being resistant only to a NAD attack, we can modify Definition \ref{def:isomo} by substituting the chase-isomorphism of item (4ii) with the KG-isomorphism, namely:
\begin{enumerate}
\item[\emph{(ii)}]  $(A[X_1],\Sigma)$ and $ (A[X_i],\Sigma)$ are KG-isomorphic through the \emph{KG-isomorphism }$\Phi_i$, for all $i\in [k]$ with $\Phi_1=id_{X_1}$.
\end{enumerate}

\textbf{Semantic utility.}
We introduce two \emph{semantic utility} metrics to quantify the coincidence of information in the original KG and in the anonymised one. This is done on a use-case basis, by focusing the metrics on a set of queries $\mc Q$ that can represent the most frequent downstream tasks asked by the users.

Notice that, given $(A,\Sigma)$ a chase-anonymisation of a KG $(G,\Sigma)$, the augmentation property (Definition \ref{def:isomo}, item (1)) implies that there is a bijective correspondence between the constants of the database $G\sub D$ and (a subset of) the constants of $A\sub D$. When evaluating a query $Q$ over $G$ and $A$, in any set operation between $Q(G)$ and $Q(A)$ we assume the equivalence of their constants under this bijective correspondence.

\begin{definition}[Semantic utility metrics]\label{def:utility}
    Let $(G,\Sigma)$ be a KG, $(A,\Sigma)$ be a $(k,x)$-chase anonymisation of it and $\mc Q = \{Q_1, \ldots, Q \sub N\}$ be a set of queries. We define the following utility metrics:
    \begin{align}
         &\mathcal{U}(G,\Sigma,A,\mc Q) := \frac{1}{N} \sum_{Q \in \mc Q}  \dfrac{ |Q(G) \cap Q(A)| }{|Q(G)|}\, ,\label{eq:utility1}\\
         &\mathcal{U}_{\bigtriangleup}(G,\Sigma,A,\mc Q) :=  \frac{1}{N} \sum_{Q \in \mc Q}  \dfrac{ |Q(G) \cap Q(A)| }{|Q(G)\cup Q(A)|}\, , \label{eq:utility_sim}
    \end{align}
with the convention that if the denominator of a term is equal to zero, then the whole fraction is set to zero. 
\end{definition}

It holds that $0\leq \mathcal{U}_{\bigtriangleup}(G,\Sigma,A,\mc Q)\leq \mathcal{U}(G,\Sigma,A,\mc Q) \leq 1$. Eq.(\ref{eq:utility1}) measures the amount of retained information of the anonymised KG with respect to the original one, i.e. the ratio of correct answers provided by $A$ with respect to the ones given by $G$. Eq.(\ref{eq:utility_sim}) also considers possible incorrect answers provided by $A$ and not existent in $G$ due to the augmentation, measuring the ratio of correct answers returned by $A$ (the ones coinciding with the ones provided by $G$) with respect to the total answers returned. In both cases, $1$ is the most desirable value, where we have complete retention of information, i.e. no loss ($\mathcal{U}$) nor incorrect answers ($\mathcal{U}_{\bigtriangleup}$). 
Example of queries on  which the semantic utility metrics can be defined are in Table \ref{tab:rules}; for their meaning, we refer the reader to Section \ref{sec:experiments}.

The utility metrics will be used by the algorithms of the next sections for targeting the chase-anonymised KG, keeping the most information possible in terms of downstream tasks. 
This would translate into bringing $\mathcal{U}_{\bigtriangleup}$ the closest to $1$.

\section{{\klone}}\label{sec:baseline}

Algorithm \ref{alg:baseline} introduces {\klone}, our first method to obtain a $(k,x)$-chase anonymisation of a KG. In particular, the anonymised graph $A$ in output is a $(k,x)$-chase anonymisation of the input KG for every $x\in [n]$; in other words, $A$ is robust to a NAR attack (and hence also to a NAD one) regardless of its size.
The anonymised graph is constructed by: (a) noising the edge weights of the original KG while maximising the utility metric $\mathcal{U}_\bigtriangleup$;
(b) copying the KG $k$ times to guarantee the $k$ chase-isomorphisms per each induced subgraph;
(c) adding synthetic edges to reach diversity of in- and out-degrees; (d)
anonymising the vertex labels (e) assign the weights to the synthetic edges while again maximising $\mathcal{U}_\bigtriangleup$.

\begin{algorithm}[t]
\footnotesize
\KwIn{A knowledge graph $(G,\Sigma)$ with $G=(V,E,L,\omega, l)$, $k\in \mathbb{N}$, set of queries $\mc Q$, 
in-degree distribution $p^-$, out-degree distribution $p^+$, $L\sub A$ set such that $L\sub A\cap L=\emptyset$, 
weight distribution $p_{\omega}$, $M\in \mathbb{N}$.}
\KwOut{A $(k,x)$-chase anonymisation $(A,\Sigma)$ of $(G,\Sigma)$ for any $x$.}
\SetKw{KwBy}{by}
\SetKwData{anonymised}{anonymised}
\SetKwData{True}{true}
\SetKwData{False}{false}
\SetKwData{UndirectedG}{$G_{\text{undirected}}$}
\SetKwData{SK}{$SK$}
\SetKwData{DSK}{$DSK$}
\SetKwData{B}{$B$}
\SetKwData{A}{$A$}
\SetKwData{VerticesToSplit}{vertices to split}

$G\gets \text{WeightNoising}(G,G,\Sigma,E,\mc Q,p_{\omega}, M)$; 

\lFor{$j=1,\dots ,k$}{$G^j\gets $ copy of $G$, $V(G^j)\gets \{v^j_1,\dots ,v^j_n\}$}
$A \gets \bigcup_{j=1}^{k}G^j$; 
 $\mathcal{A} \gets E(A) $,$\mathcal{V} \gets \{ \}$\; 
add a random edge between $G^{j}$ and $G^{j+1}$ for $j\in [k-1]$\;
    \For{$i=1,\dots ,n$}{
    $\delta_1^{-}\gets d^{-}_A(v^1_i)$;
    $\delta_1^{+}\gets d^{+}_A(v^1_i)$;
    $\mathcal{V} \gets \mathcal{V} \cup \{v^1_i\}$  \;
    \For{$j=2,\dots ,k$}{
    \For{$\varphi \in \{-,+\} $}{
    $\delta_j^{\varphi}\gets d^{\varphi}_A(v^j_i)$\;
        \While{$\exists \delta\in \{\delta^{\varphi}_1,\dots ,\delta^{\varphi}_{j-1}\}$ s.t. $\delta^{\varphi}_j=\delta$}{$\delta_j^{\varphi}\gets \max\{\delta_j^{\varphi}+1,z\sim p^{\varphi}\}$}
         $\Delta\gets \delta_j^{\varphi}-d^{\varphi}_A(v^j_i)$\; 
        $\mathcal{C} \gets \{ v\in V(A) : v\notin (N^{\varphi}_A(v^j_i)\cup \mathcal{V}\cup V(G_j))\}$\;
        \If{$|\mathcal{C}| < \Delta$}{ 
                $\mathcal{X} \gets$ set of $\Delta-|\mathcal{C}| $ new vertices\; 
                $V(A) \gets V(A)\cup \mathcal{X} $;
                $\mathcal{C} \gets \mathcal{C}\cup \mathcal{X}$\; 
            }  
            select $\Delta$ elements $\{c_1,\dots ,c_{\Delta} \}$ from $\mathcal{C}$\;
        \lIf{$\varphi=-$}{$E(A)  \gets \bigcup_{t=1}^{\Delta} ( c_t, v_i^j) \cup E(A)$}
        \lIf{$\varphi=+$}{$E(A) \gets \bigcup_{t=1}^{\Delta} (v_i^j, c_t) \cup E(A)$} 
        }
        $\mathcal{V} \gets \mathcal{V} \cup \{v^j_i\}$ \;
     }
     }
$ S\gets E(A)\setminus \mathcal{A}$; assign to each $v\in V(A)$ a different label of $L\sub A$\;
$A\gets \text{WeightNoising}(A,G,\Sigma,S, \mc Q,p_{\omega}, M)$; 
\caption{{\klone}}\label{alg:baseline}
\end{algorithm}

\begin{algorithm}[t]
\footnotesize
\KwIn{$A$, $G$ weighted labeled directed graphs, inference rules $\Sigma $, $S\subseteq E(A)$, set of queries $\mc Q$,  weight distribution $p_{\omega}$, $M\in \mathbb{N}$.}
\KwOut{A weighted labeled directed graph $A$.}
\SetKw{KwBy}{by}
\SetKwData{anonymised}{anonymised}
\SetKwData{True}{true}
\SetKwData{False}{false}
\SetKwData{UndirectedG}{$G_{\text{undirected}}$}
\SetKwData{SK}{$SK$}
\SetKwData{DSK}{$DSK$}
\SetKwData{B}{$B$}
\SetKwData{A}{$A$}
\SetKwData{VerticesToSplit}{vertices to split}
\For{$i=1,\dots ,M$}{
$A_i\gets $ copy of $A$;
\lFor{$e\in S$}{
$\omega_{A_i}(e)\gets z\sim p_{\omega}$
}
$u(i)\gets \mathcal{U}_{\Delta}(G,\Sigma,A_i,\mc Q)$
}
$I\gets \argmax_i u(i)  $; $A\gets A_{I}$;
\caption{WeightNoising}\label{alg:weight_assignement}
\end{algorithm}

More in details, (a) is performed in \textbf{line 1}, where the weights are chosen in such a way that the utility $\mathcal{U}_{\Delta}$ is maximised. 
Ideally, we would like the change of weights to completely preserve the output of the queries on the KG. 
This is done by the \emph{WeightNoising} function in Algorithm \ref{alg:weight_assignement}, which randomly samples new weights for $M$ times and chooses the ones that reach the highest value of $\mathcal{U}_{\Delta}$.
Phase (b) is in \textbf{lines 2-3}. At this point, the graph $A$ is the disjoint union of $k$ copies of $G$, which guarantees that each induced subgraph is chase-isomorphic to other $k-1$ different subgraphs according to Definition \ref{def:chase_isomorphism}. These copies are linked together in \textbf{line 4} to guarantee that $A$ is weakly connected.
Phase (c) is addressed in \textbf{lines 5--20}. The \textbf{for} loop in \textbf{line 5} iterates on the vertices of a copy of $G$ while the \textbf{for} loop in \textbf{line 7} iterates on the copies of $G$, such that $v_i^j$ refers to the $i$-th vertex in the $j$-th copy $G^j$ of $G$; the last \textbf{for} loop in \textbf{line 8} repeats the procedure both for in-degree and out-degree. The goal is to add synthetic edges such that all the copies $\{v_i^2, \dots ,v_i^k \}$ of the vertex $v_i^1$ have different sensitive attributes (Definition \ref{def:privacy}). To do so, we randomly assign a new in- (out-) degree to each of them with the use of the input degree distributions $p^-$ and $p^+$, making sure that this value is bigger than the original one (since we do not want to remove any of the original edges) and that they are all different from one another (\textbf{lines 9--11}). We then select a set of candidate vertices $\mathcal{C
}$ from (to) which adds $\Delta$ new edges to (from) each $v_i^j$ in order to meet the chosen in- (out-) degree. This set $\mathcal{C
}$ consists of all the vertices of the graph that (i) do not belong to the same copy $G^j$ of $v_i^j$, (ii) for which there is not already an existing edge and (iii) that do not belong to $\mathcal{V}$, i.e., the set of vertices for which we have already added the synthetic edges to meet the chosen in- (out-) degree (\textbf{line 13}). Condition (i) is necessary to not break the KG-isomorphisms between the subgraphs on the different copies of $G$.  
If the cardinality of $\mathcal{C}$ is less than the required number of edges that have to be added, we augment the graph with the needed number of new vertices (\textbf{lines 14--16}). We then randomly select $\Delta$ vertices from $\mathcal{C}$ and add the corresponding edges (\textbf{lines 17--19}). 
Finally, phase (d) is addressed in \textbf{line 21}, where to each vertex is assigned a new label; in phase (e) we assign a weight to each added synthetic edge (\textbf{line 22}) to maximise the utility $\mathcal{U}_{\Delta}$ again by the use of the function \emph{WeightNoising}.

\textbf{Correctness and complexity.} 
{\klone} correctly returns a $(k,x)$-chase anonymisation of the input KG and has polynomial time complexity. To argue for these properties,  
we assume that distributions $p^-$ and $p^+$ have support in $[n]$; this is reasonable as $n$ is the largest in-/out-degree that a vertex can have in $G$. Proofs are in the supplementary material~\cite{appendix}. 
\begin{lemma}\label{lem:new_nodes}
If the degree distributions $p^-$ and $p^+$ have support in $[n]$, then the anonymised graph $A$ output of Algorithm \ref{alg:baseline} is such that $kn\leq |V(A)|\leq 2kn+1$.
\end{lemma}
\begin{proof}
By \textbf{line 3} of Algorithm \ref{alg:baseline} it is clear that $|V(A)|\geq kn$, as every copy $G^j$ has cardinality equal to $n$ and no vertex is ever removed by the algorithm. To prove the other inequality, we need to show that the set $W=V(A)\setminus \bigcup_{i,j}v_i^j$, i.e. the set of new vertices that have been added to the graph in each iteration of \textbf{lines 14-16}, has at most $kn+1$ elements. Let $ D\sub {\text{out}}$ be the maximum out-degree that has been assigned to a vertex $v_i^j$ for $i\in [n]$ and $j\in [k]$, and similarly $ D\sub {\text{in}}$ for the in-degree. The key observation is that $|W|\leq D=\max\{D\sub {\text{in}}, D\sub {\text{out}}\} $. Indeed by adding $D$ new vertices, each $v_i^j$ could link to (a subset of) these vertices to meet their prescribed in- and out-degrees. It then suffices to prove that $D\leq kn+1$. 
The assigned out-degree of $v_i^j$ depends on its current out-degree $d\sub A^{\text{out}}(v_i^j)$ in the graph $A$ and on the value $z$ drawn from the distribution $p\sub{{\text{out}}}$ (\textbf{line 11}). The maximum value that $z$ can attain is $n$ by hypothesis. 
Notice that at each iteration $i$ in \textbf{line 5} and $j$ in \textbf{line 7}, the only edges that are added to the graph are the ones to/from $v_i^j$ (\textbf{lines 18-19}). In adding them, no multiple edges are allowed, as we are excluding from the set of candidates $\mathcal{C}$ of $v_i^j$ its in-/out-neighbours (\textbf{line 13}). This implies that $d\sub A^{\text{out}}(v_i^j)$ can be at most $M(i,j)=n+(i-1)k+j-i$, since $n$ is the maximum out-degree that $v_i^j$ can have within $G^j$ and $(i-1)(k-1)+j-1=(i-1)k+j-i $ is the number of vertices not in $V(G^j)$ that the algorithm have already explored in the past iterations (the cardinality of $\mathcal{V}$). Since $M(i,j)\geq n$, we have that $\max\{d\sub A^{\text{out}}(v_i^j)+1,z\sim p\sub {\text{out}}\} \leq M(i,j)+1=n+i(k-1)+j-k+1 $. 
Notice that $M(i,j)$ is strictly increasing both in $i$ and $j$, thus reaching the maximum value of $kn+1$ for $i=n$ and $j=k$, corresponding to the vertex $v_n^k$. Moreover, for fixed $i$, it assumes all different values as $j$ varies in $[k]$, as required in the \textbf{while} loop in \textbf{lines 10-11}. Therefore we have that $D\sub {\text{out}}\leq kn +1$. The same reasoning holds for $ D\sub {\text{in}}$, thus we have that $kn+1\geq D\geq |W|$.
\end{proof}
\begin{proposition}\label{prop:klone_comptime}
Algorithm \ref{alg:baseline} returns a $(k,x)$-chase anonymisation of the input knowledge graph for every $x\in [n]$. Moreover, it runs in $O(Mk^2n^2)$ time, under the hypothesis that the computation of the utility metric $\mathcal{U}_{\Delta}$ is $O(1)$ and the distributions $p^-$ and $p^+$ have support in $[n]$. 
\end{proposition}

\begin{proof}
\emph{Correctness.}
We need to show that each item of Definition \ref{def:isomo} is fulfilled by the output $A$ of the algorithm, for every $x\in [n]$. The graph $G$ is copied $k$ times in $A$ (weights included) in \textbf{line 3} and no vertex or edge is ever deleted by the algorithm, so item (1) of Definition \ref{def:isomo} is fulfilled, as well as item (2) by \textbf{line 21} and item (3) by  \textbf{line 1}. 
Let $G^1,\dots ,G^k$ be the copies of $G$ with $V(G^j)=\{v^j_1,\dots, v^j_n\}$ for $j\in [k]$.
Consider the function $\Phi_j:V(G^1)\to V(G^j)$ such that $\Phi_j(v^1_i)=v^j_i$ for each $i\in [n]$. 
Given $x\in [n]$ and $X_1$ a set of $x$ vertices of $G^1$ such that $ G^1[X_1]$ is weakly connected, the function $\Phi_{j_{|_{X_1}}}:X_1\to \Phi_j(X_1)  \subseteq V(G^j) $ is a KG-isomorphism
between $(A[X_1],\Sigma)$ and $(A[\Phi_j(X_1)],\Sigma)$. Indeed the algorithm does not add edges between vertices of the same copy of $G$ (\textbf{line 13}), so it does not modify the topology of the induced subgraphs, meeting conditions (\ref{eq:iso1}) and (\ref{eq:iso2}) of KG-isomorphism.
Moreover, by construction $X_1\cap \Phi_j(X_1)=\emptyset$, thus items (4i) and (4ii) are satisfied. We are left with verifying item (4iii). The \textbf{while} loop in \textbf{lines 10-11} guarantees that, given $i\in [n]$, to each vertex in the set $\{v^1_i,\dots ,v^k_i\}$ is associated a different out-degree and a different in-degree. This implies that $\xi\sub A(\Phi_j(v^j_i))\neq \xi\sub A(\Phi_{j'}(v^{j'}_i)) $ for all $j\neq j'$, thus satisfying item (4iii).\\
\emph{Computational time.} The computational time of the function \emph{WeightNoising} (Algorithm \ref{alg:weight_assignement}) is $O(M|S|)$, as the weight of each element of $S$, the edge set in input, is updated $M$ times. Therefore \textbf{line 1} of Algorithm \ref{alg:baseline} runs in $O(Mn^2)$ since $|E|\leq n^2$. As for \textbf{line 22}, let $V(A)=\bigcup_j V(G_j)\cup W$; by the proof of Lemma \ref{lem:new_nodes} we have that $|W|\leq kn+1$. We have already observed that Algorithm \ref{alg:baseline} (\textbf{lines 13--19}) adds edges only between $V(G_j)$ and $V(G_{j'})$ with $j\neq j'$, i.e. between vertices that belong to two different copies of $G$, and between $\bigcup_j V(G_j) $ and $W$. Consequently, in the first case it can add a maximum of $(k-1)kn^2 $ edges, while in the second case it can add up to $ 2kn(kn+1)$ edges. Therefore we have that $|S|\leq (k-1)kn^2+ 2kn(kn+1)$ and so the computational time of the function \emph{WeightNoising} in \textbf{line 22} is $O(Mk^2n^2)$. 
Let's now focus on the remaining parts of the algorithm. \textbf{Lines 2--4} clearly have a run time of $O(kn)$. 
We claim that, for fixed $i$ and $j$, the maximum number of iterations that the \textbf{while} loop in \textbf{lines 10--11} can do is $n+k$. Indeed suppose that $d\sub A^{\varphi}(v_i^j)$ is equal to some $\delta_{j'}^{\varphi}$ for $j'<j$, so that we enter the \textbf{while} loop. Let $\{\delta^{\varphi}_1,\dots , \delta^{\varphi}_{j-1}\}$ be the already assigned degrees; we have three cases:

\begin{itemize}
    \item[(i)] $ \delta^{\varphi}_s> n$ for all $s\in [j-1] $;
    \item[(ii)] $ \delta^{\varphi}_s\leq n$ for all $s\in [j-1] $;
    \item[(iii)] there exist $s,s'\in [j-1]$ such that $ \delta^{\varphi}_s> n$ and $ \delta^{\varphi}_{s'}\leq n$.
\end{itemize}

If (i) holds, $\max\{\delta_j^{\varphi}+1,z\sim p_{\varphi}\}=\delta_j^{\varphi}+1 $; therefore, since at every iteration $ \delta_j^{\varphi}$ is increased by $1$, in at most $j-2$ iterations we exit the \textbf{while} loop. 
If (ii) holds, the worst case is achieved when $z=n$ in each draw from the distribution $p_{\varphi} $ and $ \delta^{\varphi}_s= n$ for some $s\in [j-1]$. This implies that $\max\{\delta_j^{\varphi}+1,z\sim p_{\varphi}\}=z $ until $ \delta_j^{\varphi}>n$, which is achieved in at most $n-1$ iterations.
 Finally for case (iii), if $ d\sub A^{\varphi}(v_i^j)>n$ we conclude as in case (i), where the number of iterations is upper-bounded by $j-2$. If $ d\sub A^{\varphi}(v_i^j)\leq n$, the worst case is achieved when $d\sub A^{\varphi}(v_i^j)=1= \delta_1^{\varphi}$,  $z=n=\delta_2^{\varphi}$ in each draw from the distribution $p_{\varphi} $ and $ \delta_{s}^{\varphi}=n+s-2$ for each $s=3,\dots ,j-1$. In this case the \textbf{while} loop must do $n+j-3$ iterations before exiting, i.e.\ when $ \delta_j^{\varphi}$ reaches the value $n+j-3$. Since $j\leq k$, we conclude that \textbf{lines 10-11} iterate at most $n+k$ times.
Finally, by Lemma \ref{lem:new_nodes} we have that $|\mathcal{C}|\leq 2kn+1$ in \textbf{line 13}. By the same Lemma,  the total number of vertices added by the algorithm is at most $kn+1$ and so the total number of edges added is $O(k^2n^2) $, which describe the overall computational time of \textbf{lines 14--19}. Therefore the \textbf{for} loop in \textbf{lines 5--20} runs in $O(2nk(n+k+2kn+1))+O(k^2n^2)=O(k^2n^2)$. Since
\textbf{line 21} runs in $O(kn)$, the total computational time is $ O(Mk^2n^2)$.
\end{proof}

\section{{\kguard}}\label{sec:kguard}

Algorithm \ref{alg:algo_main} introduces \emph{{\kguard}}, our second proposed algorithm to obtain a $(k,x)$-chase anonymisation of a KG. Our aim is to reduce the number of synthetic vertices and edges added by {\klone} by leveraging the subgraph chase-isomorphisms that already exist in the original KG. 
The anonymised graph is constructed by:
(a) noising the edge weights of the original KG while maximising the utility metric $\mathcal{U}_\bigtriangleup$ as not to lose information;
(b) finding all the induced subgraphs of size $x$ and bucketing them into chase-isomorphic classes;
(c) selecting $k$ subgraphs in each bucket to guarantee the $k$ chase-isomorphisms and/or duplicating some of such subgraphs if the bucket has less than $k$ elements;
(d) assigning different in- and out-degrees to the vertices mapped into each other by the chase-isomorphisms to reach diversity;
(e) adding the synthetic edges to meet the requested in- and out-degrees; 
(f) anonymising the vertex labels; 
(g) assigning the weights to the synthetic edges while optimising $\mathcal{U}_\bigtriangleup$ as to maximise the usefulness of the anonymisation in terms of retained information.
Items (a), (f) and (g) are the same as in {\klone}, so we refer 
the reader to the previous section; they are in {\kguard} respectively in \textbf{line 1}, \textbf{line 31} and \textbf{line 32}. We now provide details of all the other items. 
We use the convention that if $\varphi\in \{-,+\}$, then $\mp \varphi $ denotes the opposite sign of $\varphi$.
\begin{algorithm}[t]
\footnotesize
\KwIn{A knowledge graph $(G,\Sigma)$ with $G=(V,E,L, \omega, l)$, $k,x\in \mathbb{N}$, set of queries $Q$, 
in-degree distribution $p^-$, out-degree distribution $p^+$, $L_A$ set such that $L_A\cap L=\emptyset$, weight distribution $p_{\omega}$, $M\in \mathbb{N}$.
 }
\KwOut{A $(k,x)$-chase anonymisation $(A,\Sigma)$ of $(G,\Sigma)$.}

\SetKw{KwBy}{by}
\SetKwData{anonymised}{anonymised}
\SetKwData{True}{true}
\SetKwData{False}{false}
\SetKwData{UndirectedG}{$G_{\text{undirected}}$}
\SetKwData{SK}{$SK$}
\SetKwData{DSK}{$DSK$}
\SetKwData{B}{$B$}
\SetKwData{A}{$A$}
\SetKwData{VerticesToSplit}{vertices to split}

$A\gets \text{WeightNoising}(G,G,\Sigma,E,Q,p_{\omega}, M)$\; 

  $\mathcal{H}\leftarrow\textit{ConnectedInducedSubgraphs}(A,x)$;
   $\mathbb{G}$ $\leftarrow \{ \mc G(H,\Sigma) \ | H \in \mathcal{H} \}$\; 
   
  $(\mathbb{B},\mathbb{I}) \leftarrow$ $\textit{IsomorphismBucketing}(\mathcal{H},\mathbb{G})$;
$\mathcal{V} \gets\{\}$\;
\For{$\mathcal{B} \in \mathbb{B}$}{
    choose $\mathcal{\hat{B}}\subseteq \mathcal{B}$ s.t.\ $\forall H_1\neq H_2\in \mathcal{\hat{B}}$, $V(H_1)\cap V(H_2)=\emptyset $\;
    \While{$|\mathcal{\hat{B}}| < k$}{
        choose $H\in \mathcal{\hat{B}}$; $ H' \gets \text{copy of} \ \ H $\;
        $A \gets A \cup H'$; \ \ $ \mathcal{\hat{B}} \gets \mathcal{\hat{B}} \cup H'$;
    }
   $\mathcal{B} \gets$  choose $k$ elements from $ \mathcal{\hat{B}} $; $\mathcal{V} \gets \bigcup_{H\in \mathcal{B}} V(H)\cup \mathcal{V}$;   
 }
 \lIf{$\neg IsWeaklyConnected(A)$}{add randomly an edge from $A[V]$ to each other weakly connected component}

$\{V_1,\dots ,V_s\}\gets \textit{IsomorphismPartitioning}(\mathcal{V},\mathbb{I})$ \;
\For{$i=1,\dots ,s$}{
$\{v_i^1,\dots, v_i^{n_i}\}\gets$ vertices of $V_i$\;
\For{$\varphi\in \{-,+\}$}{
$\{D^{\varphi}(v^1_i), \dots , D^{\varphi}(v^{n_i}_i)\} \! \gets \!$ \textit{ChooseDeg}$(A,\varphi,p^{\varphi},\{v_i^1,\dots, v_i^{n_i}\})$
 }}
 \For{$\varphi\in \{-,+\}$}{
\For{$v \!\in\! \mathcal{V}$}{
    $\Delta \gets D^{\varphi}(v) - d^{\varphi}\sub A(v)$; $\mathcal{X}_1\gets \{\}$; $\mathcal{X}_2\gets \{\}$\;
     $\mathcal{C} \gets  \{u\in V(A) : D^{\mp\varphi}(u) - d\sub A^{\mp\varphi}(u) >0\}$\;
     $\mathcal{C}\gets \mathcal{C} \setminus (\bigcup_{\mathcal{B}\in\mathbb{B}}\bigcup_{H\in\mathcal{B}:v\in V(H)}V(H)\cup N^{\varphi}\sub A(v)) $\; 
     \If{$|\mathcal{C}| < \Delta $}{ 
     $ m \gets \min\{\Delta-|\mathcal{C}|, |V(A)\setminus \mathcal{V}| \} $\;
     $\mathcal{X}_1\gets $ select $ m $ elements from $V(A)\setminus \mathcal{V} $\;
     \If{$|\mathcal{C}|+m < \Delta $}{
    $\mathcal{X}_2 \gets$ set of $\Delta-|\mathcal{C}|-m$ new vertices\;
    $V(A) \gets V(A)\cup\mathcal{X}_2 $; $\mathcal{C} \gets \mathcal{C}\cup \mathcal{X}_2$; 
     }
     $\mathcal{C} \gets \mathcal{C} \cup\mathcal{X}_1$; 
     }
       select $\Delta$ elements $\{c_1,\dots ,c_{\Delta} \}$ from $\mathcal{C}$\;
        \lIf{$\varphi=-$}{$E(A)  \gets \bigcup_{t=1}^{\Delta} ( c_t, v) \cup E(A)$}
        \lIf{$\varphi=+$}{$E(A) \gets \bigcup_{t=1}^{\Delta} (v, c_t) \cup E(A)$}   
}
}
assign to each $v\in V(A)$ a different label of $L_A$\;
  $A\gets \text{WeightNoising}(A,G,\Sigma,S, Q,p_{\omega}, M)$; 
\caption{{\kguard}}\label{alg:algo_main}
\end{algorithm}
\begin{algorithm}[t]
\footnotesize
\KwIn{A graph $A$, $\varphi\in \{-,+\}$, 
degree distribution $p$, a set of vertices $\{v^1,\dots ,v^m\}\subseteq V(A)$.}
\KwOut{A sequence  $\{D^{\varphi}(v^1), \dots , D^{\varphi}(v^{m})\}$ of all different degrees.}
\SetKw{KwBy}{by}
\SetKwData{anonymised}{anonymised}
\SetKwData{True}{true}
\SetKwData{False}{false}
\SetKwData{UndirectedG}{$G_{\text{undirected}}$}
\SetKwData{SK}{$SK$}
\SetKwData{DSK}{$DSK$}
\SetKwData{B}{$B$}
\SetKwData{A}{$A$}
\SetKwData{VerticesToSplit}{vertices to split}
$D^{\varphi}(v^1)\gets d^{\varphi}_A(v^1)$\;
\For{$j=2,\dots, m$}{    
    $D^{\varphi}(v^j)\gets d^{\varphi}_A(v^j)$\;
    \While{$\exists D\in \{D^{\varphi}(v^1),\dots ,D^{\varphi}(v^{j-1})\}$ s.t. $D^{\varphi}(v^{j})=D$}{$D^{\varphi}(v^j)\gets \max\{D^{\varphi}(v^j)+1,z\sim p\}$}
   } 
\caption{\textit{ChooseDeg}}\label{alg:choose_degree}
\end{algorithm}

\textit{(b) Subgraphs identification and isomorphism bucketing.} 
We first find all the weakly connected subgraphs of the KG induced by a set of $x$ vertices (\textbf{line 2}, function \emph{ConnectedInducedSubgraphs}, where we adapt the algorithm by S.\ Karakashian et al.~\cite{karakashian2013algorithm} to the case of directed graphs) and gather them in $\mathcal{H}$.
Then, for each subgraph $H \in \mathcal{H}$, we apply the reasoning rules $\Sigma$ to obtain their chase graph; together, they form the set $\mathbb{G}$ (\textbf{line 2}). 
Then, we group together the subgraphs of $\mathcal{H}$ into chase-isomorphic clusters, referred to as \emph{buckets}:  each bucket only contains subgraphs that are chase-isomorphic to each other according to Definition \ref{def:isomorphism}. This happens in \textbf{line 3}, where the function \emph{IsomorphismBucketing} returns $\mathbb{B}$, the set of buckets, and $\mathbb{I}$ the set of chase-isomorphism functions $\Psi$ between elements of $\mathbb{G}$; to do so we use a variant of the algorithm in \cite{cordella2001improved}. 
The process of isomorphism bucketing is quite straightforward, and it is done iteratively on the elements of $\mathcal{H}$. Indeed, let \( \mathbb{B} \) be the set of chase-isomorphic buckets of the first $i$ elements of $\mathcal{H}$ and let $H_{i+1}\in \mathcal{H}$ be the next considered subgraph. Since the chase-isomorphism induces an equivalence relation, it suffices to check whether $H_{i+1}$ is isomorphic to a representative of each equivalence class (bucket): if so, then $H_{i+1}$ is added to such bucket, otherwise a new bucket is created with $H_{i+1}$ as representative. 

\textit{(c) k-isomorphism.}
From each bucket $\mathcal{B}\in \mathbb{B}$, we want to choose $k$ chase-isomorphic subgraphs with a pairwise disjoint set of vertices (\textbf{lines 4--9}). 
If the bucket $\mathcal{B}$ has less than $k$ vertex-disjoint subgraphs, we copy (and add to $A$) some of them until reaching $k$ vertex-disjoint subgraphs (\textbf{lines 6--8}). 
After this process, we leave in each bucket only the $k$ chosen subgraphs (\textbf{line 9}): they will be the ones fulfilling item (4) of Definition \ref{def:isomo} of $(k,x)$-chase anonymisation. We call $\mathcal{V}$ the set of all the vertices appearing in at least a subgraph of a bucket: these are the vertices for which we want to guarantee the diversity of the sensitive attributes $\xi$.
In \textbf{line 10}, we ensure that the new graph $A$ is weakly connected by randomly adding a synthetic edge between the original graph and each other weakly connected component.
At the end of this part, each subgraph of the input KG $G$ of cardinality $x$ has other $k-1$ subgraphs in $A$ that are chase-isomorphic to it. 

\textit{(d) Diversity: degree assignment.}
\textbf{Lines 11--15} regard the assignment of a new in- and out-degree to the vertices of the chase-isomorphic subgraphs in order to guarantee diversity. To do so we first partition the set $\mathcal{V}$ in equivalence classes $\{V_1,\dots ,V_s\}$ where $u,v\in \mathcal{V}$ belong to the same class if and only if there exists a chase-isomorphism $\Psi\in\mathbb{I}$ such that $\Phi(u)=v$. In other words, all the vertices in a class must have different sensitive attributes because they belong to subgraphs that are chase-isomorphic. This is done in \textbf{line 11} by the function \emph{IsomorphismPartitioning}.
Then \textbf{lines 12--15} assign a different in- and out-degree to each vertex belonging to the same class, with the use of the function \emph{ChooseDeg}. This function, described by Algorithm \ref{alg:choose_degree}, is similar to what is done in {\klone}: we draw them at random according to the input distributions $p^{-}$ and $p^+$, while making sure that the chosen values are always bigger or equal than the original
ones and that they are all different between each others.

\textit{(e) Diversity: addition of synthetic edges.}
\textbf{Lines 16--30} add the necessary synthetic edges to meet the in- and out- degrees assigned in the previous step, thus making $A$ satisfying condition (4iii) of Definition \ref{def:isomo}.
For each $v\in \mathcal{V}$, we aim at selecting a set $\mathcal{C}$ of candidate vertices from (to) which add $\Delta$ new edges to (from) $v$ in order to achieve the chosen in-(out-) degree. Such set $\mathcal{C}$ 
is made of all the vertices $u$ of the graph such that (i) its current out-(in-) degree $d\sub A^{\mp \varphi}(u) $ is less than the assigned one $D^{\mp \varphi}(u) $, so that they have `space' for new synthetic out- (in-)coming edges (\textbf{line 19}), (ii) it does not belong to any of the subgraphs $H$ appearing in the buckets for which $v\in V(H)$ (\textbf{line 20})
and (iii) there does not already exist an edge between $v$ and $u$ (\textbf{line 20}). Item (i) is important for not exceeding $D^{\mp \varphi}(u) $, while item (ii) is crucial for not `breaking' the isomorphism between subgraphs since if $u,v\in V(H)$, then adding an edge between $v$ and $u$ would modify the topology of $H$ and hence its chase graph.
If the cardinality of $\mathcal{C}$ is less than the required number $\Delta$ of edges to be added (\textbf{line 21}), we add to it $m=\Delta-|\mathcal{C}|$ vertices of $A$ that do not belong to $\mathcal{V}$, as for them no prescribed in-/out-degree is required (\textbf{lines 22-23}). If this is not possible because $V(A)\setminus \mathcal{V}$ is too small, we create and augment $A$ with new vertices (\textbf{lines 24--26}) to be added to $\mathcal{C}$ so that $|\mathcal{C}|=\Delta$. We then randomly select $\Delta$ vertices from $\mathcal{C}$ (\textbf{line 28}) and add the corresponding edges from (to) them to (from) $v$ (\textbf{lines 29-30}).

\textbf{Correctness and complexity.}

{\kguard} is correct and has exponential worst-case complexity. The proof of the following proposition can be found in the supplementary material \cite{appendix}.
\begin{proposition}\label{prop:kguard}
Algorithm \ref{alg:algo_main} returns a $(k,x)$-chase anonymisation $(A,\Sigma)$ of the input knowledge graph $(G,\Sigma)$, for chosen $x\in [n]$ and $k\in \mathbb{N}$, $k\leq \binom{n}{x}$. 
\end{proposition}

\begin{proof} (\emph{Sketch})
We need to show that each item of Definition \ref{def:isomo} is fulfilled by the output $A$ of the algorithm. 
The first observation is that no vertex or edge is ever deleted, so item (1) is fulfilled. Item (2) is satisfied in \textbf{line 31} and item (3) by  \textbf{line 1} with the \emph{WeightNoising} function. 
\textbf{Lines 2--9} guarantee that for each induced subgraph of size $x$ there exist in $A$ other $k-1$ vertex-disjoint induced subgraphs that are KG-isomorphic to it. Indeed the \emph{IsomorphismBucketing} function group together all the subgraphs of size $x$ in KG-isomorphic classes. The $k$ subgraphs are then chosen in each class if there are enough, otherwise some subgraphs of the class are copied and added to the graph until there are $k$ of them (\textbf{lines 6--8}). Since the chosen subgraphs are all vertex-disjoint item (4i) is fulfilled. Notice also that no edge is ever added within vertices of the same subgraph (\textbf{line 20}), so the topology of the subgraphs are not modified: hence by construction item (4ii) is met. Finally, the function \emph{IsomorphingPartitioning} ensures to group together the vertices that need to have different sensitive attributes $\xi$, as it puts in the same class all the vertices for which there is a KG-isomorphism that maps one to another. Subsequently the function \emph{ChooseDeg} allocate different in- and out- degree to each vertex of a class and \textbf{lines 16--30} add the needed synthetic edges, thus satisfying item (4iii).
\end{proof}

The number of connected subgraphs with $x$ vertices can be exponentially large in $x$, specifically $ \binom{n}{x} $, potentially leading to exponential time and space complexities for the algorithm in the worst case. This issue is further compounded by the isomorphism~\cite{cordella2001improved}, as the problem of determining whether two graphs are isomorphic is still not known to be polynomial or not.\footnote{Contrarily to the subgraph isomorphism problem, known to be NP-hard. This is not our case, as we need only to assess whether two graphs are isomorphic.} Thus, in the general case, the algorithm for graph isomorphism might require exponential time. Nonetheless, it is important to note that anonymisation is a one-time process used for data release. In the subsequent experimental section, we empirically evaluate the computational time of {\kguard} and show its applicability in a wide range of scenarios: high values of $k$ and $x$ increase in general the computational cost of the anonymisation, but also implies stronger protection in terms of probability of correct re-identification ($\leq 1/k$) and maximum size of the attack we can resist ($\leq x$). The choice of $k$ and $x$ hence strongly depends on the considered KG and the needed level of privacy and it is to be evaluated case by case, based on the application and on the desired level of privacy.

\textbf{NAD attack.} In the case where we want to protect the KG only from NAD attacks and thus achieve a weaker KG-anonymisation (see Section \ref{sec:kx_iso}), it suffices to modify Algorithm \ref{alg:algo_main} in the following way.
\textbf{Line 2}: $\mathbb{G}$ $\leftarrow \{ \Sigma(H) \ | H \in \mathcal{H} \}$, we just consider the derived edges.
\textbf{Line 3}: the function   $(\mathbb{B},\mathbb{I}) \leftarrow\textit{IsomorphismBucketing}(\mathcal{H},\mathbb{G})$ returns as $\mathbb{I}$ the set of KG-isomorphisms found between the elements of $\mathbb{G}$.\\
The computational time in the general case remains the same.

\section{Experiments}\label{sec:experiments}
We evaluate {\klone} and {\kguard} for different KGs and reasoning tasks. 

We also compare our work against existing privacy preserving techniques for structural and neighbourhood attacks, stemming from the classical $k$-anonymity \cite{samarati2001protecting}. 

\textbf{Datasets.}
Our first set of experiments considers two well-known random graph models: the Erdős-Rényi and the Scale-Free network. For the Erdős-Rényi graph, we consider the directed model $D(n,M)$, where $n$ is the number of vertices and $M$ is the number of directed edges that are sampled uniformly at random among the $n^2$ possible ones. We set $M = n\ln(n) / 2$, which corresponds to the threshold for weak connectivity in the limit $n\to +\infty$ (it follows from Theorems 1 and 4 in~\cite{pike2008}). 
In this model, the in- and out-degree distribution is binomial (Poisson in the limit), thus making vertices with very large out-degree, called \emph{hubs}, unlikely to appear. The edge weights are set to $0$ with probability $0.5$, while with probability $0.5$ are sampled uniformly in $(0,1]$.
For the Scale-Free model, we consider a random graph where the out-degree profile of each vertex is randomly sampled from a truncated power-law distribution, i.e., for each vertex $v$: 
\begin{equation}\label{eq:power_law}
 \mathbb{P}( d\sub{G}^{+}(v)=d) = \dfrac{d^{-\alpha}}{{\sum_{k=0}^{n-1}k^{-\alpha}}}\,\,\, \text{ if } 0\le d< n,\,\,\,  =0\,\,\, \text{otherwise,}
\end{equation}
where $\alpha>0$ is a parameter. Low values of $\alpha$ correspond to a high probability of having nodes with large out-degree, while for high values of $\alpha$, almost all the vertices have small out-degree with high probability.
A feature of graphs with such power law distributions (particularly when $\alpha$ is small) is the presence of \emph{hubs}. These networks are called \emph{scale free} \cite{barabasi2008,bollobas2003} and have been shown to model economic networks \cite{GARLASCHELLI2005}. 
The edge weights are assigned by sampling in $[0,1]$ uniformly at random, and then rescaled such that for each $v \in V$, it holds that
 $ \sum_{u \in N^{-}(v)}  \omega(u,v)  \leq 1$:
this is because we want to model a company network with ownership relationships between companies, where the total sum of shares owned of a node cannot be more than $100\%$.
For both models, we enforce the generated graph to be weakly connected by randomly adding edges between its components. Alternatively, our algorithms can be applied separately to each component.

Lastly, we consider five different real-world graphs from the literature, namely MovieLens small~\cite{harper2015movielens}, Econ-Mahindas~\cite{nr}, Power-1138-Bus~\cite{nr}, Bitcoin Alpha~\cite{kumar2018rev2} and an anonymised fragment of the Company Ownership graph~\cite{magnanimi2023reactive}. This latter graph and its importance among central banks and financial authorities have already been discussed in the introduction and Figure~\ref{fig:knowledge_graph}. 
From each of the above-mentioned networks, we extract and anonymise a weakly connected component, whose characteristics can be found in Table~\ref{tab:real_world_data}.

\textbf{Reasoning Tasks.}
Our experiments use two distinct sets of reasoning rules. For the networks modelling company ownerships (namely the \textit{Scale-Free} and the \textit{Company Ownership} graphs), we use the \emph{control} rules between two companies ($\Sigma = \{ \sigma_1, \sigma_2 \}$ in Table \ref{tab:rules}). 
For all the others, we use rules on \emph{reachability}: a derived edge from vertex $u$ to vertex $v$ is added if there exists a path from $u$ to $v$ whose product of the edges' weight is greater than 0. In Vadalog~\cite{bellomarini2018vadalog}, this can be written as the set of rules $\Sigma=\{\sigma_3, \sigma_4\}$ in Table \ref{tab:rules}.
\textbf{Evaluation Metrics.} 
We evaluate our work according to three key aspects of synthetic graph data, namely \emph{fidelity}, \emph{utility}, and \emph{privacy}~\cite{potluru2023synthetic}. 

The fidelity measures how much the anonymised graph is statistically \emph{close} to the original one. We evaluate it in terms of the number of vertices added to the original graph and of the Wasserstein-1 distance~\cite{dobrushin} of the degree and weight distributions between the original and the anonymised graph. 

We then evaluate the utility of the anonymised graph with the metrics $\mathcal{U}$ and $\mathcal{U}_{\bigtriangleup}$ of Definition~\ref{def:utility}, measuring the correctness of potential down-stream tasks in the form of reasoning queries. 
We consider the queries of Table \ref{tab:rules}.
Query $Q_1$ returns all the couple of nodes connected by an edge of weight $>0.3$, 
$Q_2$ returns each couple of nodes that are connected in both directions with different weights, 
$Q_3$ selects all the vertices of the KG with at least $2$ out-going edges, while $Q_4$ selects all the vertices of the KG with at least $2$ out-going edges with weight greater than $q$, 
$Q_5$ returns each couple for which there is a direct control (shares above $50\%$), $Q_6$ selects all the couple for which one controls the other with a (ownership) path of length three, and $Q_7$ returns all the couple $(x,y)$ such that $x$ controls $y$ and $x$ is not controlled by any entity, thus being the ultimate controller of $y$. Queries $\{Q_1,Q_2,Q_3,Q_4\}$ can be applied to any generated network described in the previous section, while queries $\{Q_5,Q_6,Q_7\}$ are specifically for the company ownership networks.
For $Q_4$ we set $q=0$ for all the considered networks but the company ownership ones, for which we set $q=0.5$, as it is the threshold for establishing control.

The set of queries used to evaluate the utility metrics will be a subset $\mathcal{Q}$ of $\{Q_1,Q_2,Q_3,Q_4,Q_5,Q_6,Q_7\}$, chosen to best represent the potential analysis tasks based on the network type, i.e., company ownership networks or general graphs.

Finally, we evaluate the improvement in terms of privacy reached by our approach (the weaker one based on KG-isomorphism) with respect to classical structural anonymisation. For the comparison, we consider an algorithm 

that provides anonymisation by forming $k$ isomorphic subgraphs in the graph (\emph{$k$-isomorphism}) and we measure the percentage of subgraphs that are correctly anonymised by this approach, meaning that we count how many isomorphic copies are also KG-isomorphic. Our experiments confirm that neglecting derived knowledge leads to potential leaks of information as classical approaches are not able to provide privacy for each substructure: the presence of derived edges might break the isomorphism, making a subgraph uniquely identifiable.

\textbf{Implementation.}

We implemented our analyses and algorithms in Python, using NetworkX~\cite{hagberg2008exploring}. 
In the experiments we model the in-degree and out-degree distributions, namely $p^-$ and $p^+$, as a binomial distribution fit on the degree profile of the input graphs. 

The weight distribution $p_{\omega}$ is chosen by partitioning the weights of the input graph into 200 equal-width bins, from which we estimate an empirical distribution. 

We recall that different approaches, like a non-parametric KDE~\cite{davis2011remarks}, can be easily integrated. Finally, the results report the mean and the standard error over five different random trials; while the $\downarrow$ ($\uparrow$) symbol alongside each metric indicates that a lower (higher) value is better.

\subsection{Simulated and real-world networks}
In this section, we extensively evaluate our approaches on different graph models, varying the number of vertices, edges, and privacy requirements, such as the number of chase-isomorphic subgraphs $k$ and the size of considered subgraphs $x$. We refer to {\kguard} and 
{\textsc{kguard (KG-ISO)}} for the chase-isomorphism and the weaker KG-isomorphism anonymisation (see Section \ref{sec:kx_iso}), respectively. Where not otherwise stated, we consider $n=500$, $k=3$ and $x=4$. 

\begin{figure}[t]
    \centering
    \begin{overpic}[width=0.32\linewidth]{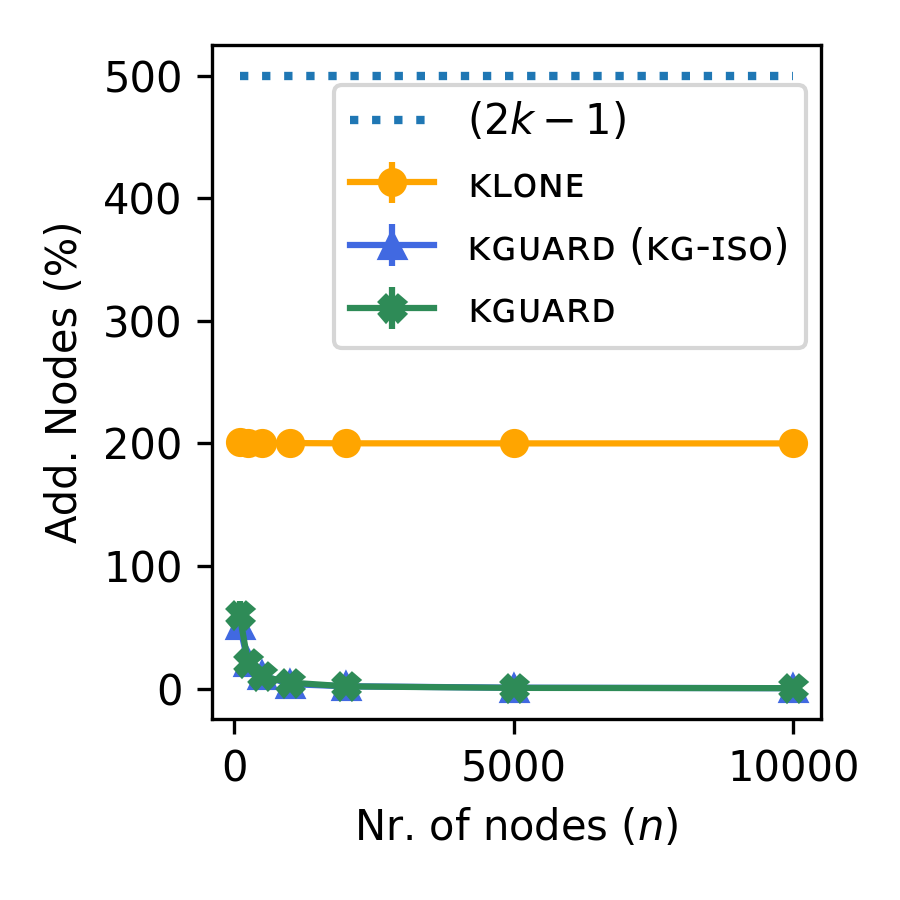}
        \put(5,5){ (a)}
    \end{overpic}
    \begin{overpic}[width=0.32\linewidth]{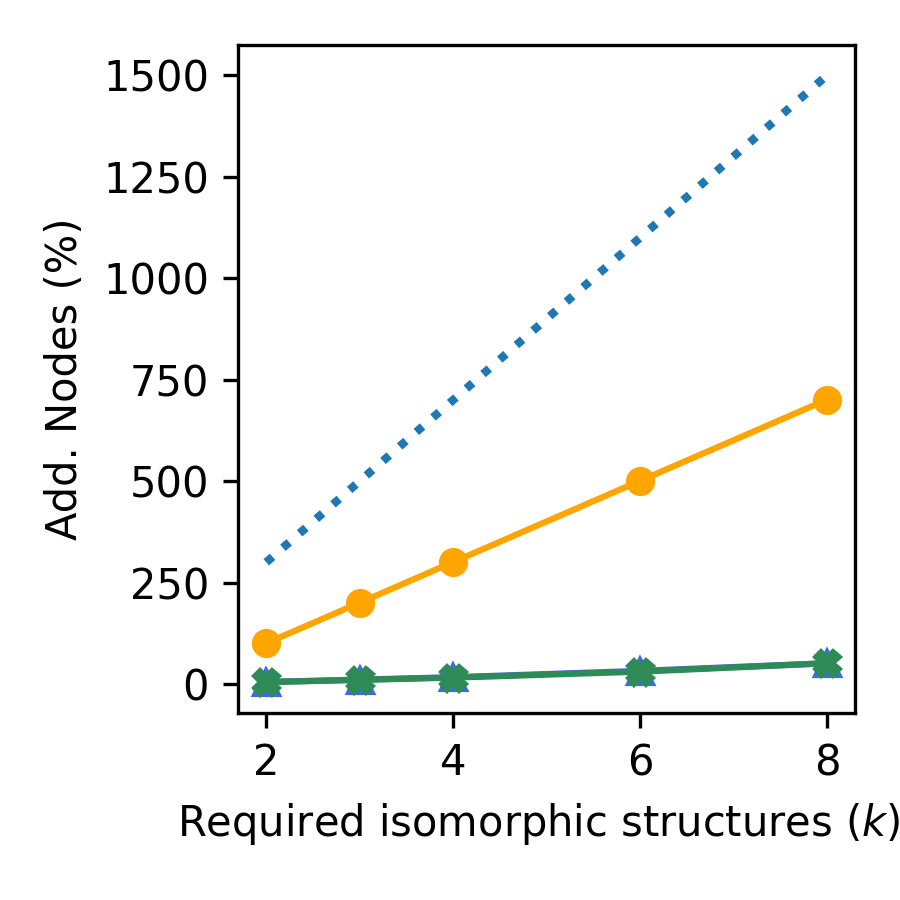}
        \put(-1,5){ (c)}
    \end{overpic}
    \begin{overpic}[width=0.32\linewidth]{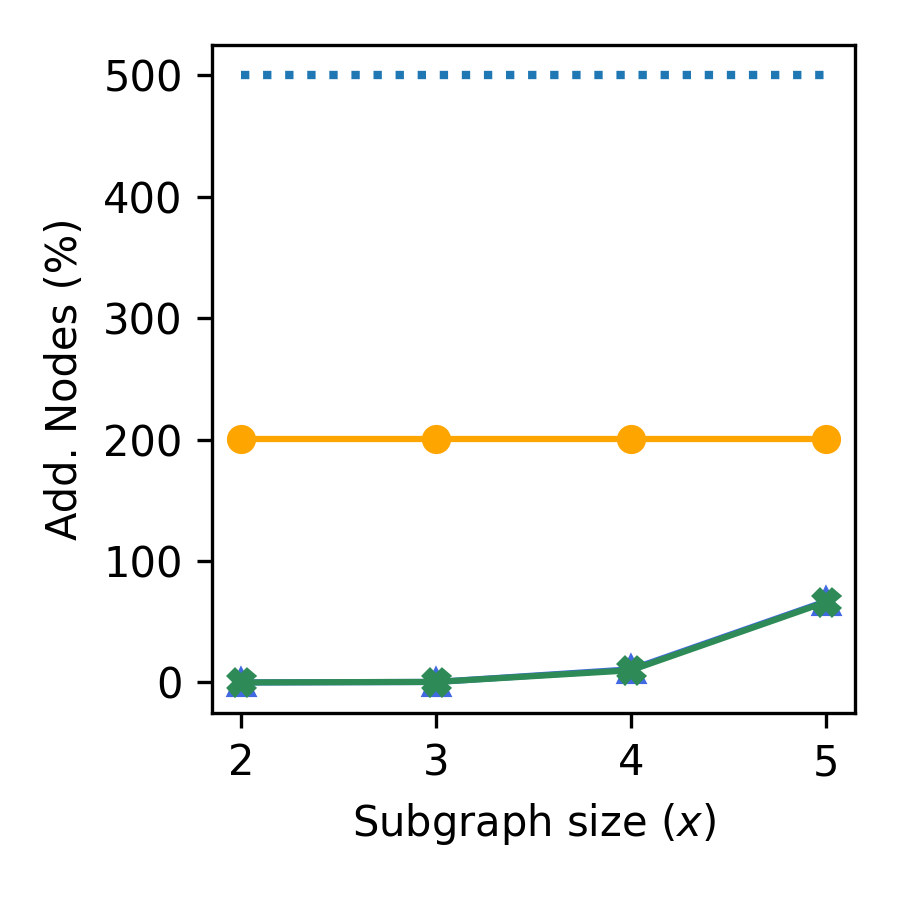}
        \put(5,5){ (e)}
    \end{overpic}
    \vfill
    \begin{overpic}[width=0.32\linewidth]{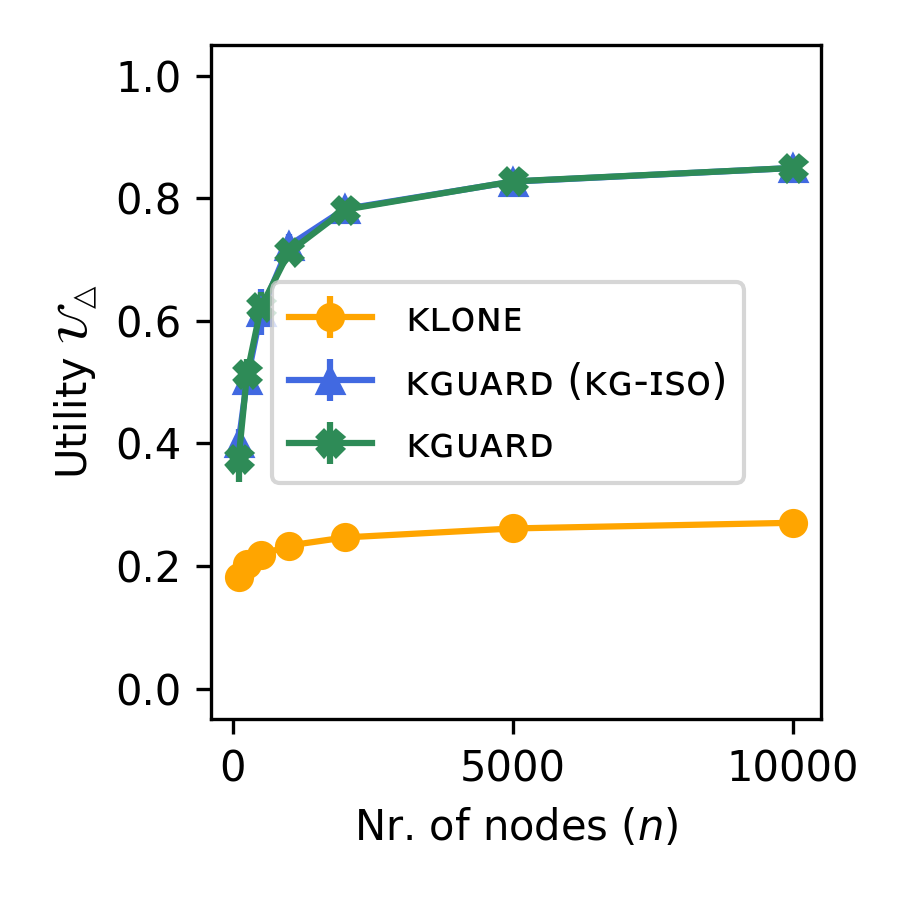}
        \put(5,5){ (b)}
    \end{overpic}
    \begin{overpic}[width=0.32\linewidth]{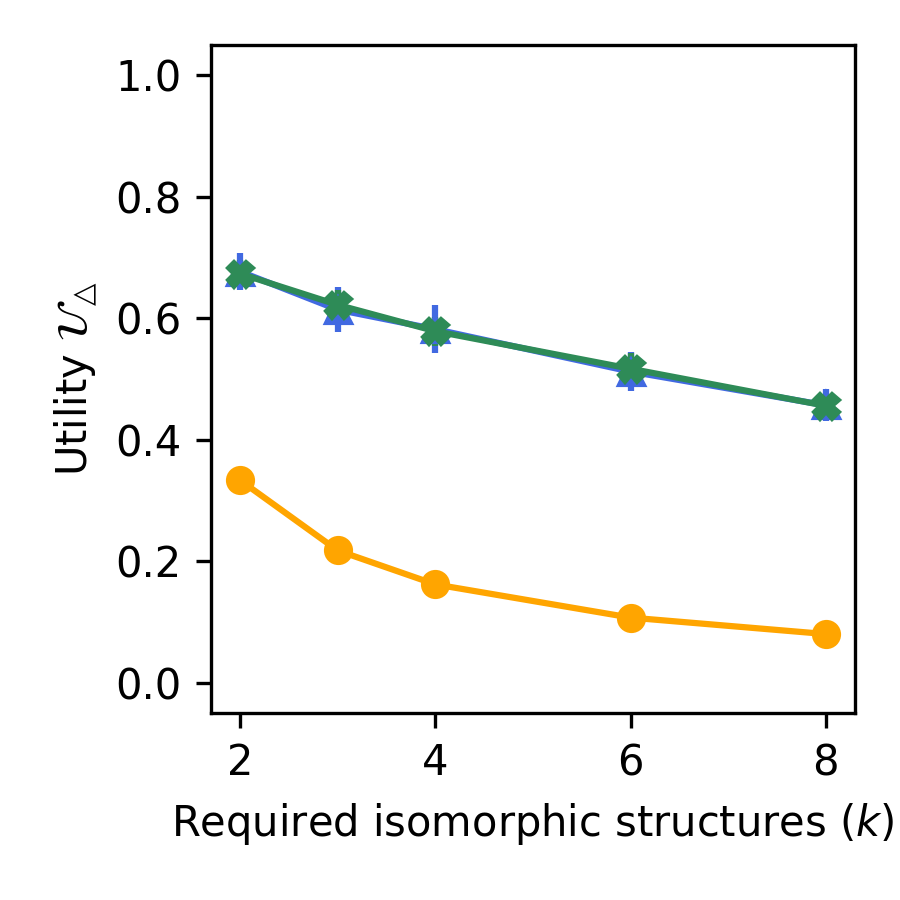}
        \put(-1,5){ (d)}
    \end{overpic}
    \begin{overpic}[width=0.32\linewidth]{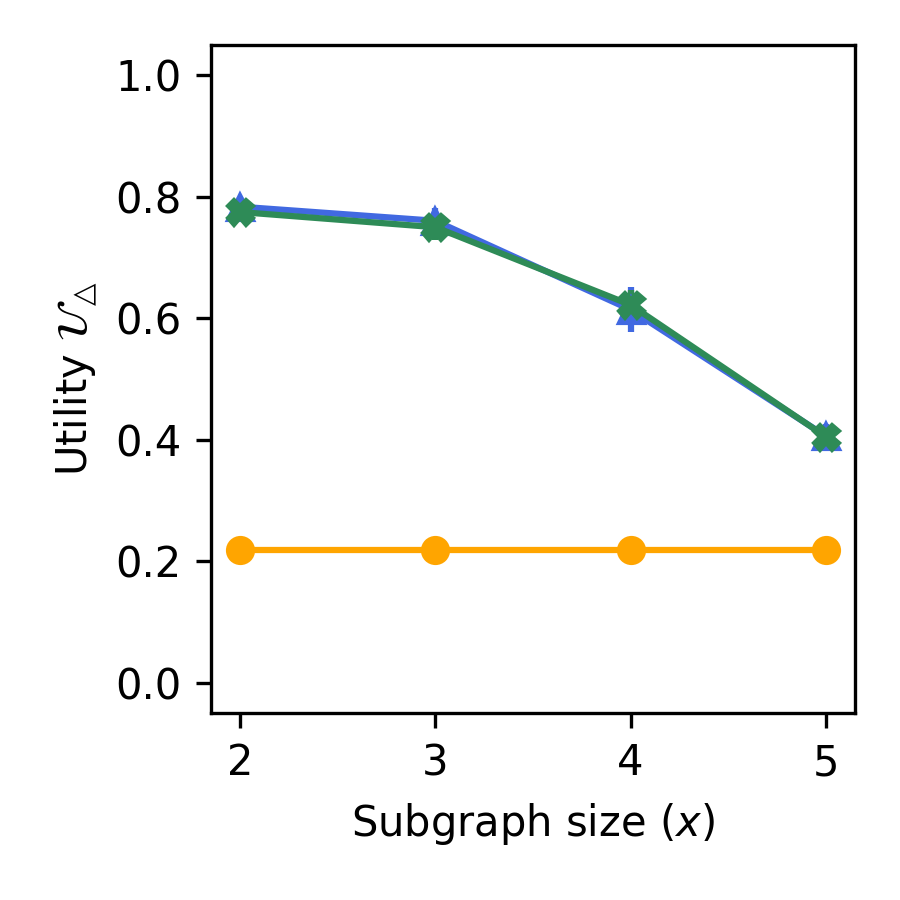}
        \put(5,5){ (f)}
    \end{overpic}
    \captionof{figure}{Erdős-Rényi model with $\Sigma = \{ \sigma_3, \sigma_4 \}$ and $\mathcal{Q}=\{Q_5, Q_6 \}$, varying $n$, $k$ and $x$.}
    \label{fig:vary_plot_erdos}
\end{figure}

\textbf{Erdős-Rényi.}
In Figure~\ref{fig:vary_plot_erdos}, we investigate the fidelity and utility of the anonymisation for the Erdős-Rényi model. In Figure (\ref{fig:vary_plot_erdos}a) and (\ref{fig:vary_plot_erdos}b) we vary the number of vertices $n$ of the input graph from 100 to 10,000. Figure (\ref{fig:vary_plot_erdos}a) shows the percentage of vertices added during the anonymisation by the algorithms  (\emph{Add. Nodes (\%)}): {\klone} consistently adds around 200\% more vertices as expected, influenced by the privacy requirement $k=3$, whereas {\kguard} introduces fewer additional vertices by leveraging existing isomorphic subgraphs within the input graph. The dotted horizontal line represents the theoretical upper bound of $(2k -1)\%$ added vertices for {\klone} (Lemma \ref{lem:new_nodes}). The efficiency of {\kguard} with respect to {\klone}, in terms of added structures, is also reflected in Figure (\ref{fig:vary_plot_erdos}b), where {\kguard} exhibits significantly higher utility compared to {\klone}. We also notice that {\textsc{kguard (KG-ISO)}} has slightly better performance than {\kguard}, as fewer redundant structures are required to satisfy the weaker anonymisation requirement. In terms of computational cost, both algorithms require approximately the same amount of time to anonymize a graph (e.g., around 3 hours for a graph with $5,000$ nodes). However, for larger graphs, {\klone} proves to be more efficient, requiring only 5 hours for a graph with 10,000 nodes while {\kguard} and \textsc{kguard (KG-ISO)} take nearly 10 hours.

In Figures (\ref{fig:vary_plot_erdos}c) and (\ref{fig:vary_plot_erdos}d), we vary the privacy requirement $k$ from 2 to 8. As expected, in Figure (\ref{fig:vary_plot_erdos}c) {\klone} shows a linear increase in additional vertices with respect to $k$, while {\kguard} exhibits just a slight increase. This trend is also reflected in the utility (Figure \ref{fig:vary_plot_erdos}d), where {\klone} has a much lower utility compared to {\kguard}. In Figures (\ref{fig:vary_plot_erdos}e) and (\ref{fig:vary_plot_erdos}f), we evaluate the anonymisation quality by varying the size $x$ of the considered subgraphs. The performance of {\klone} remains unaffected as expected, since the algorithm provides a $(k,x)$-chase anonymisation for any $x\in [n]$ (Proposition \ref{prop:klone_comptime}), instead {\kguard} overhead increases with the size of subgraphs $x$. In general, {\kguard} shows a better performance, while {\klone} might be the better choice when $x$ is sufficiently large or if we are uncertain about the attacker knowledge. In fact, while {\kguard} overhead and computational time increase with $x$, {\klone} maintains a consistent cost as a single anonymisation $A$ of $G$ is effective for any choice of $x$. Notice that the utility $\mathcal{U}$ is close to 1 (the highest value) for all settings and algorithms.

\begin{table}[t]
    \resizebox{1\linewidth}{!}{
    \begin{tabular}{lllrrrrr}
    \toprule
$n$ & $\alpha$ & Algo & $\mathcal{U}$ $\uparrow$ & $\mathcal{U}_{\bigtriangleup}$ $\uparrow$ & Weights $\downarrow$ & Degree $\downarrow$ & A.Nodes(\%)$\downarrow$\\
\midrule
500 & 3 & {\klone} & \bf  0.96 ± 0.01 & 0.62 ± 0.00 & 0.10 ± 0.00 & 3.43 ± 0.09 & 200.2 ± 0.22 \\
    &  & {\kguard} & \bf  0.96 ± 0.01 & \bf  0.75 ± 0.02 & \bf 0.05 ± 0.00 & \bf 2.14 ± 0.11 & \bf 32.7 ± 16.33 \\
500 &  4 & {\klone} & \bf  0.97 ± 0.01 & 0.61 ± 0.00 & 0.10 ± 0.01 & 3.20 ± 0.06 & 200.3 ± 0.26 \\
    &  & {\kguard} & \bf  0.97 ± 0.01 & \bf  0.77 ± 0.03 & \bf 0.05 ± 0.01 & \bf 2.10 ± 0.05 & \bf 15.65 ± 13.5 \\
500 & 5  & {\klone} & \bf 0.96 ± 0.01 & 0.61 ± 0.00 & 0.10 ± 0.01 & 3.05 ± 0.05 & 200.0 ± 0.32 \\
& & {\kguard} & \bf 0.96 ± 0.01 & \bf 0.76 ± 0.02 & \bf  0.05 ± 0.00 & \bf  1.89 ± 0.13 & \bf  15.49 ± 5.55 \\
  2k & 5 & {\klone} & \bf  0.97 ± 0.01 & 0.61 ± 0.00 & 0.10 ± 0.00 & 3.10 ± 0.05 & 200.0 ± 0.04 \\
&  & {\kguard} & 0.96 ± 0.00 &\bf  0.82 ± 0.01 & \bf  0.04 ± 0.00 & \bf  0.75 ± 0.05 & \bf  1.77 ± 0.67 \\
  5k & 5 & {\klone} & \bf  0.93 ± 0.05 & 0.61 ± 0.00 & 0.13 ± 0.03 & 3.15 ± 0.02 & 200.0 ± 0.02 \\
&& {\kguard} & 0.92 ± 0.05 & \bf  0.81 ± 0.03 & \bf 0.03 ± 0.02 & \bf 0.35 ± 0.03 & \bf 0.65 ± 0.54 \\
  10k &5  & {\klone} & \bf 0.95 ± 0.04 & 0.61 ± 0.00 & 0.11 ± 0.03 & 3.17 ± 0.04 & 200.0 ± 0.01 \\
&& {\kguard} & 0.93 ± 0.05 & \bf  0.82 ± 0.03 & \bf  0.04 ± 0.03 & \bf 0.18 ± 0.01 & \bf 0.25 ± 0.18 \\
    \bottomrule
    \end{tabular}
    }
    \captionof{table}{Anonymisation for scale-free model with $k=3$, $x=4$, $\Sigma = \{ \sigma_1, \sigma_2 \}$,  $\mathcal{Q}=\{Q_5\}$, and varying $n$ and $\alpha$.}
    \label{tab:erdos_power_law}
\end{table}

\textbf{Scale-free.}
Table~\ref{tab:erdos_power_law} reports the performance of our algorithms on a scale-free network with varying numbers of vertices up to $10,000$, and $\alpha \in \{3,4,5\}$. The results show that {\kguard} generally outperforms {\klone} as it adds significantly fewer nodes to the anonymised graph (\% Add. Nodes $\downarrow$), and it has distributions that are closer to the original graph both for weights (\emph{Weights} $\downarrow$), and degrees (\emph{Degree} $\downarrow$), along with better utility metrics.
Similarly to the Erdős-Rényi model, as the number of nodes increases, {\kguard} has a significantly lower overhead than {\klone}: e.g., with $10,000$ nodes {\kguard} adds only $0.25\%$ of new nodes.
Finally, the utility metrics, namely $\mathcal{U}$ and $\mathcal{U}_{\bigtriangleup}$, remain consistently high, close to the optimal value of 1. Especially for {\kguard}, this indicates that the anonymised graph retains nearly all the information of the original graph, when evaluated on the given set of queries.

Notice that, as the performance difference between {\textsc{kguard (KG-ISO)}} and {\kguard} is consistent and minimal, we omit the former in the remaining results.

\textbf{Real-world graphs.}
We investigate the performance of our approaches for five real-world graphs from various domains, including energy and economics. Table~\ref{tab:real_world_data} reports for each graph the number of vertices and edges, along with the anonymisation performance for {\klone} and {\kguard}, with $k=3$ and $x=4$. 
 
The results show that {\kguard} consistently outperforms {\klone} across almost all the evaluated metrics. As {\kguard} introduces significantly less structural overhead (\emph{Add. Nodes (\%)}), it better preserves the degree (\emph{Degree} $\downarrow$) and weight (\emph{Weights} $\downarrow$) distributions in the anonymised graph. Both {\kguard} and {\klone} consistently achieve almost a perfect utility $\mathcal{U}$, close to 1, indicating that the anonymised graph $A$ has no loss of information compared to graph $G$ in terms of utility queries. 

However, for the utility $\mathcal{U}_{\bigtriangleup}$ {\klone} shows a more significant presence of incorrect answers due to the \(k\) copies it makes of the original graph, as clearly reflected in the higher percentage of additional nodes. In contrast, {\kguard} maintains a relatively high utility, thanks to a low percentage of additional nodes, meaning that only a few vertex additions are needed to ensure the \(k\) chase-isomorphisms.

\subsection{Utility and state-of-the-art comparison}

\textbf{Utility.} Table~\ref{tab:multiple_utilities} shows the performance of our algorithms when different sets of queries are considered in the semantic utility metrics, both on a scale-free network and an Erdős-Rényi graph of $2,000$ nodes. 
For example, the first results in Table~\ref{tab:multiple_utilities} regard a scale-free network with reasoning rules $\Sigma = \{ \sigma_1, \sigma_2 \}$, where the metrics $\mathcal{U}$ and $\mathcal{U}_{\Delta}$ are computed on the query $Q_3$ asking \textit{"How many companies have direct control over a company by holding more than 50\% of its shares?"}.

For both algorithms, the analysts obtain an approximately correct lower bound on the real number (since $\mathcal{U}$ is close to 1), meaning that the companies that would have been identified in response to this query in the original graph $G$ are also present in the anonymised graph $A$. However, to preserve company identities, both approaches have created some redundant structures that might also positively answer this question. The utility $\mathcal{U}_{\bigtriangleup}$ also takes into account the number of additional companies retrieved from this query that exist in $A$ but not in $G$. In general {\kguard} has higher utility $\mathcal{U}_{\bigtriangleup}$, as it depends on the number of additional structures created in the anonymised graph $A$. For {\klone}, the additional structures only partially depend on the input graph and are mainly influenced by the number of anonymisation copies $k$ required (as shown also in Figure~\ref{fig:vary_plot_erdos}). In contrast, {\kguard} works specifically on each subgraph, and additional structures are only needed if the original graph contains very unique subgraphs that require duplication. 

This difference is also evident across the different sets of queries: while {\klone} maintains relatively low utility $\mathcal{U}_{\bigtriangleup}$, {\kguard} achieves values close to 1, representing the optimal solution.

The most challenging case for both algorithms occurs with the query set $\mathcal{Q}=\{Q_5, Q_6\}$, as only a small subset of nodes from the original graph match such queries. Therefore, we notice that even a limited number of incorrect new matches in the anonymised graph substantially impact the  metric $\mathcal{U}_{\bigtriangleup}$.

To summarise, Table~\ref{tab:multiple_utilities} shows that {\kguard} generally performs better than {\klone} due to the fewer added structures, though different query sets may lead to varying results, depending on their complexity.

\begin{table}
\caption{Anonymisation results for real-world graphs with $k=3$ and $x=4$.}\label{tab:real_world_data}
\resizebox{1\linewidth}{!}{
\begin{tabular}{llcc}
\toprule
 Dataset & Metric & {\klone}  & {\kguard} \\
\midrule

Company Ownership~\cite{magnanimi2023reactive}&utility $\mathcal{U}$ $\uparrow$ & 
\bf  {0.99 $\pm$ 0.00} & {0.99 $\pm$ 0.01} \\
$n = 3,000$, \ \ $|E| = 3,900$ &utility $\mathcal{U}_{\bigtriangleup}$ $\uparrow$ & 0.66 $\pm$ 0.01 & \bf 0.82 $\pm$ 0.01 \\
$\Sigma = \{ \sigma_1, \sigma_2 \}$, $\mathcal{Q}=\{Q_5\}$ & weights $\downarrow$ & 0.47 $\pm$ 0.00 & \bf 0.40 $\pm$ 0.00 \\
& degree $\downarrow$ & 3.78 $\pm$ 0.01 & \bf 2.55 $\pm$ 0.04 \\
& add. nodes (\%) $\downarrow$ & 200.0 $\pm$ 0.1 & \bf 40.8 $\pm$ 2.77 \\

Econ-Mahindas~\cite{nr}& utility $\mathcal{U}$ $\uparrow$ & 1.00 ± 0.00 & \bf 1.00 ± 0.00 \\
$n = 1,200$, \ \ $|E| = 3,300$& 
utility $\mathcal{U}_{\bigtriangleup}$ $\uparrow$ & 0.24 ± 0.00 & \bf 0.68 ± 0.00 \\
$\Sigma = \{ \sigma_3, \sigma_4 \}$, $\mathcal{Q}=\{Q_3, Q_4 \}$ & weights $\downarrow$ & 0.38 ± 0.00 & \bf 0.34 ± 0.00 \\
& degree $\downarrow$ & 4.54 ± 0.05 & \bf 2.33 ± 0.07 \\
 & add. nodes (\%) $\downarrow$ & 200.1 ± 0.1 & \bf 8.22 ± 0.26 \\

MovieLens small~\cite{harper2015movielens} &utility $\mathcal{U}$ $\uparrow$ & \bf 1.00 ± 0.00 & \bf 1.00 ± 0.00 \\
$n = 2,100$, \ \ $|E| = 3,200$ & 
utility $\mathcal{U}_{\bigtriangleup}$ $\uparrow$ & 0.09 ± 0.00 & \bf 0.83 ± 0.01 \\
$\Sigma = \{ \sigma_3, \sigma_4 \}$, $\mathcal{Q}=\{Q_3, Q_4 \}$ & weights $\downarrow$ & 0.42 ± 0.00 & \bf 0.21 ± 0.00 \\
& degree $\downarrow$ & 4.19 ± 0.03 & \bf 0.35 ± 0.01 \\
& add. nodes (\%) $\downarrow$ & 200.1 ± 0.1 & \bf 2.12 ± 0.08 \\

 Power-1138-Bus~\cite{nr} &  utility $\mathcal{U}$ $\uparrow$ & \bf 1.00 ± 0.00 & \bf 1.00 ± 0.00 \\
$n = 1,100$, \ \ $|E| = 2,600$ & 
utility $\mathcal{U}_{\bigtriangleup}$ $\uparrow$ & 0.25 ± 0.00 & \bf 0.90 ± 0.01 \\
$\Sigma = \{ \sigma_3, \sigma_4 \}$, $\mathcal{Q}=\{Q_3, Q_4 \}$ & weights $\downarrow$ & 0.20 ± 0.00 & \bf 0.15 ± 0.00 \\
& degree $\downarrow$ & 3.78 ± 0.07 & \bf 0.64 ± 0.05 \\
& add. nodes (\%) $\downarrow$ & 200.2 ± 0.1 & \bf 0.35 ± 0.00 \\

Bitcoin Alpha~\cite{kumar2018rev2} & utility $\mathcal{U}$ $\uparrow$ & \bf 1.00 ± 0.00 & \bf  1.00 ± 0.00 \\
$n = 1,700$, \ \ $|E| = 3,100$ 
& utility $\mathcal{U}_{\bigtriangleup}$ $\uparrow$ & 0.13 ± 0.00 & \bf 0.64 ± 0.01 \\
$\Sigma = \{ \sigma_3, \sigma_4 \}$, $\mathcal{Q}=\{Q_3, Q_4 \}$ & weights $\downarrow$ & 0.36 ± 0.00 & \bf 0.30 ± 0.00 \\
& degree $\downarrow$ & 4.12 ± 0.06 & \bf  0.85 ± 0.04 \\
& add. nodes (\%) $\downarrow$ & 200.1 ± 0.1 & \bf 6.21 ± 0.08 \\

\midrule
\end{tabular}
}
\end{table}

\begin{table}
\caption{Anonymisation results for different query sets and graph models ( $k=3$, $x=4$, $n=2,000$, $\alpha=5$).}\label{tab:multiple_utilities}
\resizebox{1\linewidth}{!}{
\begin{tabular}{lllll}
\toprule
Model & Query set & Metric & {\klone} & {\kguard} \\
\midrule

Scale-free & $\mathcal{Q}=\{Q_5\}$ & add. nodes (\%) $\downarrow$ & 200.0 ± 0.0 & \bf 1.77 ± 0.67 \\
$\Sigma = \{ \sigma_1, \sigma_2 \}$&
& utility $\mathcal{U}_{\bigtriangleup}$ $\uparrow$ & 0.61 ± 0.00 & \bf 0.82 ± 0.01 \\
& & utility $\mathcal{U}$ $\uparrow$ &  \bf 0.97 ± 0.01 &  0.96 ± 0.00 \\

Scale-free & $\mathcal{Q}=\{Q_6\}$ & add. nodes (\%) $\downarrow$ & 200.1 ± 0.0 & \bf 2.17 ± 0.25 \\
$\Sigma = \{ \sigma_1, \sigma_2 \}$&
& utility $\mathcal{U}_{\bigtriangleup}$ $\uparrow$ & 0.50 ± 0.00 & \bf 0.83 ± 0.29 \\
& & utility $\mathcal{U}$ $\uparrow$ &  \bf 1.00 ± 0.00 & \bf 1.00 ± 0.00 \\

Scale-free & $\mathcal{Q}=\{Q_3, Q_4\}$ & add. nodes (\%) $\downarrow$ & 200.1 ± 0.0 & \bf 1.86 ± 0.72 \\
$\Sigma = \{ \sigma_1, \sigma_2 \}$&
 & utility $\mathcal{U}_{\bigtriangleup}$ $\uparrow$ & 0.02 ± 0.00 & \bf 0.10 ± 0.02 \\
& & utility $\mathcal{U}$ $\uparrow$ &  \bf 0.95 ± 0.06 & 0.94 ± 0.06 \\

Scale-free & $\mathcal{Q}=\{Q_7\}$ & add. nodes (\%) $\downarrow$ & 200.1 ± 0.0 & \bf 1.87 ± 0.62 \\
$\Sigma = \{ \sigma_1, \sigma_2 \}$&
 & utility $\mathcal{U}_{\bigtriangleup}$ $\uparrow$ & 0.56 ± 0.00 & \bf 0.65 ± 0.01 \\
& & utility $\mathcal{U}$ $\uparrow$ &  0.79 ± 0.01 & \bf 0.80 ± 0.01 \\

Erdős-Rényi &  $\mathcal{Q}=\{Q_1\}$ & add. nodes (\%) $\downarrow$ & 200.1 ± 0.1 & \bf 1.84 ± 0.58 \\
$\Sigma = \{ \sigma_3, \sigma_4 \}$&
 & utility $\mathcal{U}_{\bigtriangleup}$ $\uparrow$ & 0.63 ± 0.00 & \bf 0.93 ± 0.00 \\
& & utility $\mathcal{U}$ $\uparrow$ &  \bf 1.00 ± 0.00 & 0.99 ± 0.00 \\

Erdős-Rényi &  $\mathcal{Q}=\{Q_3, Q_4 \}$ & add. nodes (\%) $\downarrow$ & 200.1 ± 0.1  & \bf 1.84 ± 0.58 \\
$\Sigma = \{ \sigma_3, \sigma_4 \}$&
 & utility $\mathcal{U}_{\bigtriangleup}$ $\uparrow$ & 0.25 ± 0.00 & \bf 0.78 ± 0.01 \\
& & utility $\mathcal{U}$ $\uparrow$ &  \bf 1.00 ± 0.00 & \bf 1.00 ± 0.00 \\

Erdős-Rényi &  $\mathcal{Q}=\{Q_2\}$ & add. nodes (\%) $\downarrow$ & 200.1 ± 0.1 & \bf 1.86 ± 0.58 \\
$\Sigma = \{ \sigma_3, \sigma_4 \}$&
 & utility $\mathcal{U}_{\bigtriangleup}$ $\uparrow$ & 0.59 ± 0.02 & \bf 0.68 ± 0.07 \\
& & utility $\mathcal{U}$ $\uparrow$ &  \bf 0.96 ± 0.06 & \bf 0.96 ± 0.06 \\
\bottomrule
\end{tabular}
}
\end{table}

\textbf{State-of-art comparison.}  We now investigate the privacy of our approaches compared to classical structural anonymisation 
(\emph{k-Iso}). This method anonymises graphs by forming $k$ pairwise isomorphic subgraphs, making them indistinguishable to an attacker. However, as we qualitatively showed in Figure~\ref{fig:ex1_reidentification}, neglecting derived links in KGs results in severe privacy issues and information leaks. In Table~\ref{tab:state_of_art_comparison}, we quantitatively evaluate such leaks by measuring the percentage of subgraph structures that are correctly anonymised, i.e. for which there exists other $k-1$ KG-isomorphic subgraphs within the KG; we call this index $\delta$-\emph{anonymity}. Results are shown for $x\!=\!4$ and $k\!=\!3$. 
The table confirms our theoretical analysis: our approaches anonymise each individual subgraph and consistently achieve $\delta$-\emph{anonymity} $= 1$. Contrarily, the state-of-the-art approach \emph{k-Iso} does not protect the privacy of all entities: in the worst case (Bitcoin-Alpha), only 60\% of subgraphs are effectively not uniquely identifiable, leaving 40\% of potentially identifiable entities by an attacker. 
The best case for the state-of-the-art algorithm is
Econ-Mahindas, still showing around $8\%$ of subgraph structures vulnerable to attacks.

\begin{table}
\caption{Anonymisation rate with $k=3$ and $x=4$.}\label{tab:state_of_art_comparison}

\resizebox{1\linewidth}{!}{
\begin{tabular}{llccc}
\toprule
\multirow{2}{*}{Reasoning} & \multirow{2}{*}{Graph} & \multicolumn{3}{c}{$\delta-anonymity$ $\uparrow$} \\
\cmidrule(lr){3-5}
 &  & K-Iso & {\klone} & {\kguard} \\  
\midrule
Reach& Scale-free ($n = 100,\alpha=5$) & 0.795 & \bf 1.000 & \bf 1.000 \\

$\Sigma = \{ \sigma_3, \sigma_4 \}$ & Scale-free ($ n = 500,\alpha=5$) & 0.873 & \bf 1.000 & \bf 1.000 \\
 &MovieLens small~\cite{harper2015movielens} & 0.818 & \bf 1.000 & \bf 1.000 \\
 &Power-1138-Bus~\cite{nr} & 0.723 & \bf 1.000 & \bf 1.000 \\
 &Bitcoin Alpha~\cite{kumar2018rev2} & 0.605  & \bf 1.000 & \bf 1.000 \\
&Econ-Mahindas~\cite{nr} & 0.923 & \bf 1.000 & \bf 1.000 \\
Control & Scale-free ($n = 100,\alpha=5$) & 0.720 & \bf 1.000 & \bf 1.000 \\
$\Sigma = \{ \sigma_1, \sigma_2 \}$ & Scale-free ($n = 500,\alpha=5$)& 0.728 & \bf 1.000 & \bf 1.000 \\
  & Company Own~\cite{magnanimi2023reactive} & 0.680 & \bf 1.000 & \bf 1.000 \\
\bottomrule
\end{tabular}
}
\end{table}

\section{Related Work}\label{sec:relwork}
Several definitions of privacy have been proposed over the years, ranging from traditional syntactic privacy definitions~\cite{samarati2001protecting} to more recent semantic ones 
like differential privacy~\cite{jian2023publishing,mueller2022sok,jiang2021applications,dwork2006differential}. 

Differential privacy (DP) frameworks are designed to protect individual entities during data analysis, i.e. they prevent an attacker from determining whether a specific individual was included in the input data.

However, DP faces significant challenges when applied to highly correlated and network data~\cite{jiang2021applications}. Adding noise to nodes or edges often fails to conceal the overall structure and relationships within the data. In addition, the anonymization algorithm must be carefully designed to ensure that the DP mechanism preserves the statistics of interest. 

These limitations make DP not the best candidate in a KG setting, where structural anonymisation is often preferred to provide privacy while preserving data utility~\cite{hoang2023protecting}.

In the last years, structural anonymisation concepts originally developed for relational databases \cite{casas2017survey,ji2016graph} have been extended to graph data, including models such as 
$k$-degree and $k$-neighbourhood anonymity~\cite{majeed2020anonymization,ren2022kt}.
These approaches 
usually focus on modifying the graph so that it exhibits at least $k$ ``similar'' (e.g., isomorphic) structures with respect to the adversary knowledge~\cite{hay2008,zhou2008}. 

However, existing methods target only specific graph types --- like directed~\cite{kiabod2021fast,casas2019k} or weighted graphs~\cite{li2016practical} --- 
and when applied to KGs they may expose sensitive information, since adversaries can exploit the rich KG attributes~\cite{hoang2020cluster}. 

Specifically for KGs, only few anonymisation solutions exist~\cite{hoang2023protecting,hoang2023time,hoang2021privacy,hoang2020cluster,thouvenot2020knowledge}. They focus mainly on the privacy of single entities~\cite{hoang2020cluster,thouvenot2020knowledge}, providing also personalized anonymisation~\cite{hoang2023protecting}; as well as the privacy protection of sequential published data~\cite{hoang2023time,hoang2021privacy}. Differently from existing work, we focus on a more complex attack model considering subgraph structures, where attackers can exploit logical reasoning and derived knowledge, commonly present in business settings, to identify target entities.

\section{Conclusion}\label{sec:conclusion}
We discussed the application of privacy protecting schemes in the realm of KGs,  
showing that existing structural approaches fail to maintain privacy in presence of derived knowledge. 
We proposed new structural anonymisation techniques that guarantee privacy also under attacks with information on the reasoning rules, and described and evaluated two algorithms achieving it by 
generating synthetic variations of the input graph while preserving its utility for downstream tasks.
\section{AI-Generated Content Acknowledgement}
Authors acknowledge that no GenAI tools were used in any stage of the research, nor in the writing.
\printbibliography
\end{document}